\documentclass[fleqn,usenatbib]{mnras}
\usepackage[T1]{fontenc}
\usepackage{ae,aecompl}
\usepackage{epsf,palatino,url,wrapfig ,fancyhdr,multirow,lscape,appendix,rotating, setspace,amsmath,amssymb}
\usepackage{graphics,epsfig,txfonts}
\usepackage{graphicx}
\usepackage{gensymb}
\usepackage{longtable}
\usepackage{amssymb}
\usepackage{xcolor}
\usepackage{color}
\usepackage{lipsum}
\usepackage{textcomp}
\usepackage{threeparttable}
\usepackage{enumerate}
\usepackage{placeins}
\usepackage{float}
\usepackage[normalem]{ulem}
\usepackage{soul}
\usepackage{blindtext}

\usepackage{tikz,hyperref}
\usepackage{academicons}

\newcommand{\msol}{\ensuremath{M_{\odot}}\xspace}

\newcommand{\rxte}{\textit{RXTE}\xspace}
\newcommand{\nustar}{\textit{NuSTAR}\xspace}

\newcommand{\nicer}{\textit{NICER}\xspace}

\newcommand{\swift}{\textit{Swift}\xspace}

\newcommand{\maxi}{\textit{MAXI}\xspace}

\newcommand{\source}{SMC~X-2\xspace}

\newcommand{\lumcgs}{erg~s$^{-1}$}

\usepackage{tikz,hyperref}
\usepackage{academicons}
\definecolor{lime}{HTML}{A6CE39}
\DeclareRobustCommand{\orcidicon}{%
	\begin{tikzpicture}
	\draw[lime, fill=lime] (0,0) 
	circle [radius=0.16] 
	node[white] {{\fontfamily{qag}\selectfont \tiny ID}};
	\draw[white, fill=white] (-0.0625,0.095) 
	circle [radius=0.007];
	\end{tikzpicture}
	\hspace{-2mm}
}

\foreach \x in {A, ..., Z}{%
	\expandafter\xdef\csname orcid\x\endcsname{\noexpand\href{https://orcid.org/\csname orcidauthor\x\endcsname}{\noexpand\orcidicon}}
}

\newcommand{\orcid}[1]{\href{https://orcid.org/#1}{\textcolor[HTML]{A6CE39}{\aiOrcid}}}

\title[SMC~X-2 with {NICER} and {NuSTAR} ]
{On the cyclotron absorption line and evidence of the spectral transition in SMC X-2 during 2022 giant outburst}
\author[Jaisawal et al.]
{G. K. Jaisawal,$^{1}$\thanks{E-mail: gaurava@space.dtu.dk} 
G. Vasilopoulos,$^{2,3}$
S. Naik,$^{4}$
C. Maitra,$^{5}$
C. Malacaria,$^{6}$
\newauthor
B. Chhotaray,$^{4,7}$
K. C. Gendreau,$^{8}$
S. Guillot,$^{9}$
M. Ng,$^{10}$
and A. Sanna$^{11}$
\\
$^1$ DTU Space, Technical University of Denmark, Elektrovej 327-328, DK-2800 Lyngby, Denmark\\ 
$^2$Universit\'e de Strasbourg, CNRS, Observatoire astronomique de Strasbourg, UMR 7550, F-67000 Strasbourg, France\\
$^3$ Department of Physics, National and Kapodistrian University of Athens, University Campus Zografos, GR 15783, Athens, Greece \\
$^{4}$Astronomy and Astrophysics Division, Physical Research Laboratory, Navrangpura, Ahmedabad - 380009, Gujarat, India\\
$^{5}$Max-Planck-Institut f{\"u}r extraterrestrische Physik, Gie{\ss}enbachstra{\ss}e 1, 85748 Garching, Germany\\
$^{6}$International Space Science Institute (ISSI), Hallerstrasse 6, 3012 Bern, Switzerland\\
$^{7}$Indian Institute of Technology Gandhinagar, Palaj, Gandhinagar - 382055, Gujarat, India\\
$^{8}$Astrophysics Science Division, NASA's Goddard Space Flight Center, Greenbelt, MD 20771, USA \\
$^{9}$ Institut de Recherche en Astrophysique et Planétologie, UPS-OMP, CNRS, CNES, 9 avenue du Colonel Roche, BP 44346, F-31028 Toulouse Cedex 4, France \\
$^{10}$ MIT Kavli Institute for Astrophysics and Space Research, Massachusetts Institute of Technology, Cambridge, MA 02139, USA\\
$^{11}$Dipartimento di Fisica, Università degli Studi di Cagliari, SP Monserrato-Sestu km 0.7, 09042 Monserrato, Italy \\}
\pubyear{2023}

\begin{document}
\label{firstpage}
\pagerange{\pageref{firstpage}--\pageref{lastpage}}
\maketitle

\begin{abstract}

We report comprehensive spectral and temporal properties of the Be/X-ray binary pulsar SMC X-2 using X-ray observations during the 2015 and 2022 outbursts. The pulse profile of the pulsar is unique and strongly luminosity dependent.  It evolves from a broad-humped into a double-peaked profile above luminosity 3$\times$10$^{38}$~ergs~s$^{-1}$. The pulse fraction of the pulsar is found to be a linear function of luminosity as well as energy. We also studied the spectral evolution of the source during the latest 2022 outburst with \emph{NICER}. The observed photon index shows a negative and positive correlation below and above the critical luminosity, respectively, suggesting evidence of spectral transition from the sub-critical to super-critical regime. The broadband spectroscopy of four sets of \emph{NuSTAR} and XRT/\emph{NICER} data from both outbursts can be described using a cutoff power-law model with a blackbody component. In addition to the 6.4 keV iron fluorescence line, an absorption-like feature is clearly detected in the spectra. The cyclotron line energy observed during the 2015 outburst is below 29.5 keV, however latest estimates in the 2022 outburst suggest a value of 31.5~keV. Moreover, an increase of 3.4~keV is detected in the cyclotron line energy at equal levels of luminosity observed in 2022 with respect to 2015. The observed cyclotron line energy variation is explored in terms of accretion induced screening mechanism or geometrical variation in line forming region. 

\end{abstract}


\section{INTRODUCTION}
\label{sec:intro}

Be/X-ray binaries (BeXRBs) represent two-thirds of the population of high mass X-ray binaries (HMXBs). These systems consist of a massive ($>10 \msol$) optical companion and a compact object (usually a neutron star) in a binary system. The optical companion in BeXRBs is a non-supergiant OB spectral type star that shows Balmer series emission lines and infrared excess at a point in its life \citep{Reig2011}. The above characteristics originate from an equatorial circumstellar disk that is formed around the Be star due to its rapid rotation at velocities of more than $75\%$ of Keplerian limit \citep{Porter2003}. The compact object in the system, on the other hand, accretes directly from the Be-circumstellar disk. Two kinds of X-ray outbursts are observed from BeXRBs.  First, Type-I outbursts are short (only a few weeks long), (quasi-)periodic events that reach a peak luminosity
of $\le$10$^{37}$\lumcgs and occur close to the periastron passage of the binary system.
The second category of outbursts is giant in nature, where the peak luminosity reaches $\ge$10$^{37-38}$\lumcgs. The latter usually lasts for a multiple or significant portion of the orbit and does not follow any orbital dependencies. 

A significant population of BeXRBs is found in the Milky Way and the Magellanic Clouds with luminosities in the range of 10$^{34-38}$\lumcgs \citep{Liu2006, Reig2011}. Our Galaxy hosts about 60--70 such systems \citep{Reig2011, Walter2015}. The Small Magellanic Cloud (SMC; a neighboring irregular dwarf galaxy), however, contains about 68 pulsars (and a total of 122 sources including candidates) despite being only a few percent of the mass of the Milky Way Galaxy \citep{Reig2011, Walter2015, Coe2015MNRAS.452..969C, Haberl2016, Yang2017ApJ...839..119Y, Vinciguerra2020}. 

SMC~X-2 is a 2.37~s pulsating  source in a BeXRB system located inside SMC \citep{Schurch2011MNRAS.412..391S} at a distance of 62 kpc \citep{Hilditch2005MNRAS.357..304H, Graczyk2014ApJ...780...59G}. It was the second brightest X-ray source in the galaxy when {\it SAS~3} discovered it in October 1977. The recorded luminosity was 8.4$\times$10$^{37}$\,ergs\,s$^{-1}$ (assuming  a distance of 65 kpc) in the 2--11 keV energy band \citep{Clark1978ApJ...221L..37C, Clark1979ApJ...227...54C}. Later observations with the {\it HEAO}, {\it Einstein} and {\it ROSAT} missions demonstrated the X-ray transient nature of the source \citep{Marshall1979ApJS...40..657M, Seward1981ApJ...243..736S, Kahabka1996A&A...312..919K}. Thanks to \rxte ({\it Rossi X-ray Timing Explorer}) and {\it ASCA} ({\it Advanced Satellite for Cosmology and Astrophysics}), coherent X-ray pulsations from SMC~X-2 were detected for the first time during the 2000 January-April outburst \citep{Corbet2001ApJ...548L..41C}. This discovery made it possible to identify the accreting object as a neutron star. The optical observations of the system, on the other hand, suggested two possible early-type stars at an angular separation of 2.5 arcsec as the potential counterparts of the pulsar. The Optical Gravitational Lensing Experiment (OGLE) I-band photometric studies found variability at a period of 18.62$\pm$0.02 d from one of the stars (northern star) \citep{Schurch2011MNRAS.412..391S}. This periodicity closely matched the orbital period of $\approx$18.4 days measured by studying the X-ray pulse-period modulation during the 2002 and 2015 outbursts \citep{Townsend2011MNRAS.416.1556T, Li2016ApJ...828...74L}. The optical counterpart (northern star) is identified to be an O9.5 III-V emission star \citep{McBride2008MNRAS.388.1198M}.      

The system displayed major outbursts during 1977 October \citep{Marshall1979ApJS...40..657M, Seward1981ApJ...243..736S}, 2002 January-April  \citep{Corbet2001ApJ...548L..41C}, 2015 August-November \citep{Kennea2015ATel.8091....1K, Jaisawal2016MNRAS.461L..97J}, and 2022 June-August \citep{Coe2022ATel15500....1C}.
Detailed studies with {\it XMM}-Newton, {\it Swift}/XRT, and {\it NuSTAR} were performed during the 2015 outburst when the pulsar was accreting in the super-Eddington regime with a peak outburst luminosity of 6$\times$10$^{38}$~ergs~s$^{-1}$ \citep{Negoro2015ATel.8088....1N, Kennea2015ATel.8091....1K, Jaisawal2016MNRAS.461L..97J}. Ionized emission lines from N, O, Ne, Si, and Fe were detected in the data from the {\it XMM}-Newton observation  \citep{Palombara2016MNRAS.458L..74L}. Moreover, with broadband coverage of \nustar  ~ in the range of 3-79 keV, the neutron star magnetic field was estimated  for the first time to be $\approx$3$\times$10$^{12}$~G based on the discovery of a cyclotron resonance scattering feature (CRSF) at around 27 keV \citep{Jaisawal2016MNRAS.461L..97J}. A negative correlation between the luminosity and cyclotron line energy was also reported in the study. In addition to this, the pulsar is known to show luminosity-dependent pulse profiles evolving from single to double-peaked \citep{Li2016ApJ...828...74L, Jaisawal2016MNRAS.461L..97J, Roy2022ApJ...936...90R}. The onset of the propeller regime observed in late 2015 also provided a measurement of the magnetic field to be $\approx$3$\times$10$^{12}$~G \citep{Lutovinov2017ApJ...834..209L}, in agreement with the estimation through CRSF detection. 

\begin{figure}
\centering
\includegraphics[height=3.9in, width=3.2in]{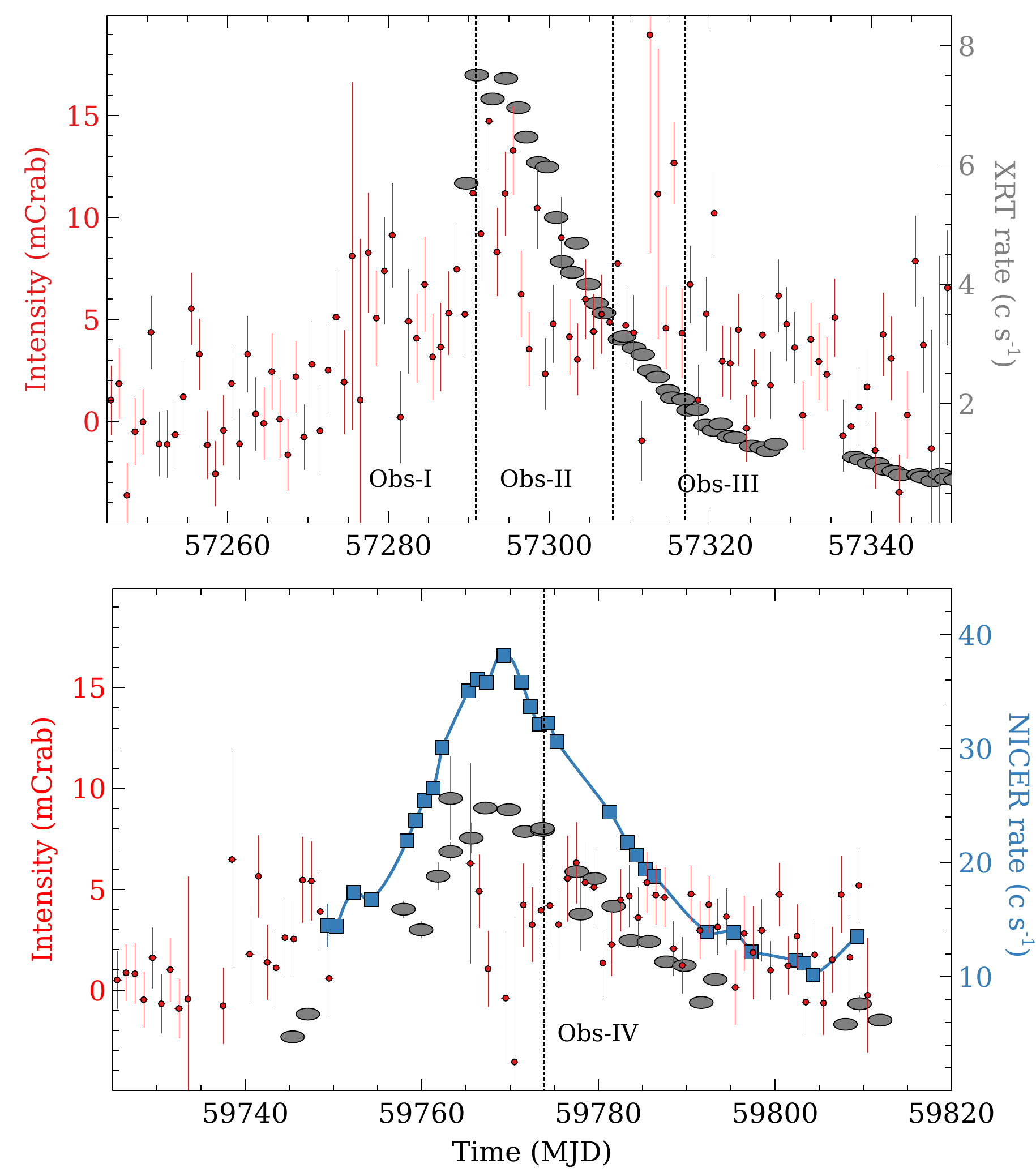} 
\caption{The light curves of SMC~X-2 during its 2015 (top) and 2022 (bottom) X-ray outbursts with {\it MAXI} (2-20 keV; red dot), \nicer (0.5-10 keV; blue square), and \swift/XRT (0.3-10 keV; grey ellipse). The vertical dotted lines show the start date of the \nustar observations. In the bottom panel, the XRT light curve is scaled by a factor of seven to compare with the \nicer~ light curve.}
\label{outburst_lc}
\end{figure}

After seven years of X-ray quiescence, SMC~X-2 became active on 2022 June \citep{Kennea2022ATel15434....1K}. The pulsar reached a peak luminosity of 1$\times$10$^{38}$~ergs~s$^{-1}$, three weeks after its detection in the soft X-ray band \citep{Coe2022ATel15500....1C}. Unlike the 2015 outburst, the \maxi/GSC \citep{Matsuoka2009PASJ...61..999M} did not detect any significant outburst peak in 2--20 keV band due to the relatively moderate nature of the outburst in addition to the observational coverage gap (see, e.g.,  Figure~\ref{outburst_lc}).  We monitored the target closely with \nicer and \nustar to understand the spectral and temporal evolution during the 2022 outburst. We also analyzed three \nustar observations from the 2015 outburst to examine the cyclotron line and its evolution and also studied the broadband temporal and spectral properties of the source.  Section~2 reports the details of the observations and data analysis methods. The timing and spectral results are presented in Section~3, followed by a discussion and conclusion in Section~4.

\section{Observations and Data Analysis}
\label{sec:data}

\subsection{\nicer }

Launched in 2017 June, the \nicer X-ray timing instrument (XTI, \citealt{Gendreau2012, Gendreau2016}) is a soft X-ray telescope (0.2--12 keV) attached to the International Space Station. The XTI consists of a total of 56 co-aligned concentrator optics. Each optics collimates the soft X-ray photons onto a silicon drift detector at the focus \citep{Prigozhin2012}. A high time resolution of $\sim$100~ns (rms) and spectral resolution  of $\approx$85~eV at 1~keV are achieved by \nicer. The effective area of the instrument is $\approx$1900~cm$^2$ at $1.5$~keV, with 52 active detectors. The total field of view of \nicer is $\approx$30~arcmin$^2$. 

\nicer monitored  \source{} after the onset of the 2022 outburst in June.  
We used a total of 37 observations under the ObsIDs 52028301xx with a net exposure time of $\approx 39.6$~ks for studying the pulsar emission during the outburst between 2022 June 20 and August 17. We reduced the \nicer data using the {\tt nicerl2} \footnote{\url{https://heasarc.gsfc.nasa.gov/docs/nicer/analysis_threads/nicerl2/}} script available under \textsc{HEASoft} version 6.30. The reprocessing is done in the presence of the gain and calibration database files of version 20200722. The cleaned events obtained from the pipeline are used in our further analysis. In the beginning, the events from  the South Atlantic Anomaly region were filtered out. We further applied standard filtering criteria based on the elevation angle from the Earth limb, pointing offset, and the offset from  the bright Earth. Good time intervals were produced using the  {\tt nimaketime} task. Final spectra and light curves were created with the \textsc{XSELECT} package of \textsc{FTOOLS}. The spectral background corresponding to each observation is generated using the {\tt nibackgen3C50}\footnote{\url{https://heasarc.gsfc.nasa.gov/docs/nicer/toolsnicer_bkg_est_tools.html}} tool \citep{Remillard2022AJ....163..130R}. The response matrix and ancillary response files of version 20200722 are applied in our spectral analysis.

 \begin{table}
\centering
\caption{Log of simultaneous observations of SMC~X-2 with {\it NuSTAR} and {\it Swift}/XRT or {\it NICER} during its 2015 and 2022 X-ray outbursts.}
\begin{tabular}{lccc}
\hline
\\
Observatory/		&ObsID   	    &Start Date   		&Exposure  \\
Instrument		   &	     	      & 			       &(ks) \\
\hline
{\it NuSTAR}		 &90102014002 	  &2015-09-25T21:51:08	  &24.5 \\ 
{\it Swift}/XRT  &00034073002   &2015-09-25T22:32:58    &1.8 \\
\\
{\it Swift}/XRT  &00081771002   &2015-10-12T21:30:58    &1.5 \\ 
{\it NuSTAR}		 &90102014004	  &2015-10-12T21:41:08	  &23 \\ 
\\
{\it Swift}/XRT  &00034073042   &2015-10-21T14:08:58    &4 \\
{\it NuSTAR}     &90101017002   &2015-10-21T21:31:08    &26.7\\

\\
{\it NICER}  &5202830119   &2022-07-13T07:50:21    &1.1 \\
{\it NuSTAR}     &90801319002   &2022-07-13T20:18:43    &42.7\\

\hline
\end{tabular}
\label{obs}
\end{table}  

\subsection{\nustar} \label{sec:nustar}
\nustar is the first hard X-ray focusing observatory launched in 2012 June  \citep{Harrison2013}. It consists of two co-aligned grazing angle incidence telescopes. The mirrors in each optic are coated with multilayers of Pt/SiC and W/Si that reflect the soft to hard X-ray photons. A Cadmium-Zinc-Telluride detector at the focal point of each unit is sensitive to the 3--79 keV photons.  Following the recent outburst of SMC~X-2, we requested a \nustar ToO observation. The target was observed for an effective exposure of 42.7 ks on 2022 July 13-14 (Table~\ref{obs}). Standard  analysis procedures were followed for data reduction with {\tt NuSTARDAS} 1.9.7 software. Unfiltered events were processed in the presence of the CALDB of version 20220802 using {\it nupipeline} task. Source products are extracted from a circular region of 120~arcsec radius around the central coordinates on each detector using the {\it nuproducts} task. The background products are also accumulated in a similar manner from a source-free circular region of 120~arcsec radius. The background-subtracted light curves from both detector modules of \nustar were combined for our timing studies. 

We also reduced the data from the \nustar observations of SMC~X-2 during its 2015 major outburst (Table~\ref{obs}) by following the above procedure. The first observation was close to the peak of the outburst in 2015 September, while the other two were in the declining phase. Corresponding \swift/XRT observations are analyzed to perform broadband spectroscopy (see also \citealt{Jaisawal2016MNRAS.461L..97J, Lutovinov2017ApJ...834..209L} for XRT data analysis and descriptions; \citealt{Evans2009}). All four \nustar observations in the 2015-2022 timeline are simply referred to as Obs-I, Obs-II, Obs-III, and Obs-IV in the paper. For spectroscopy, we have grouped each \nicer, \nustar, and XRT spectra for a minimum of 32 counts per channel bin to achieve a good signal-to-noise ratio using {\tt grppha}.

\begin{figure}
\centering
\includegraphics[height=5.1in, width=3.2in]{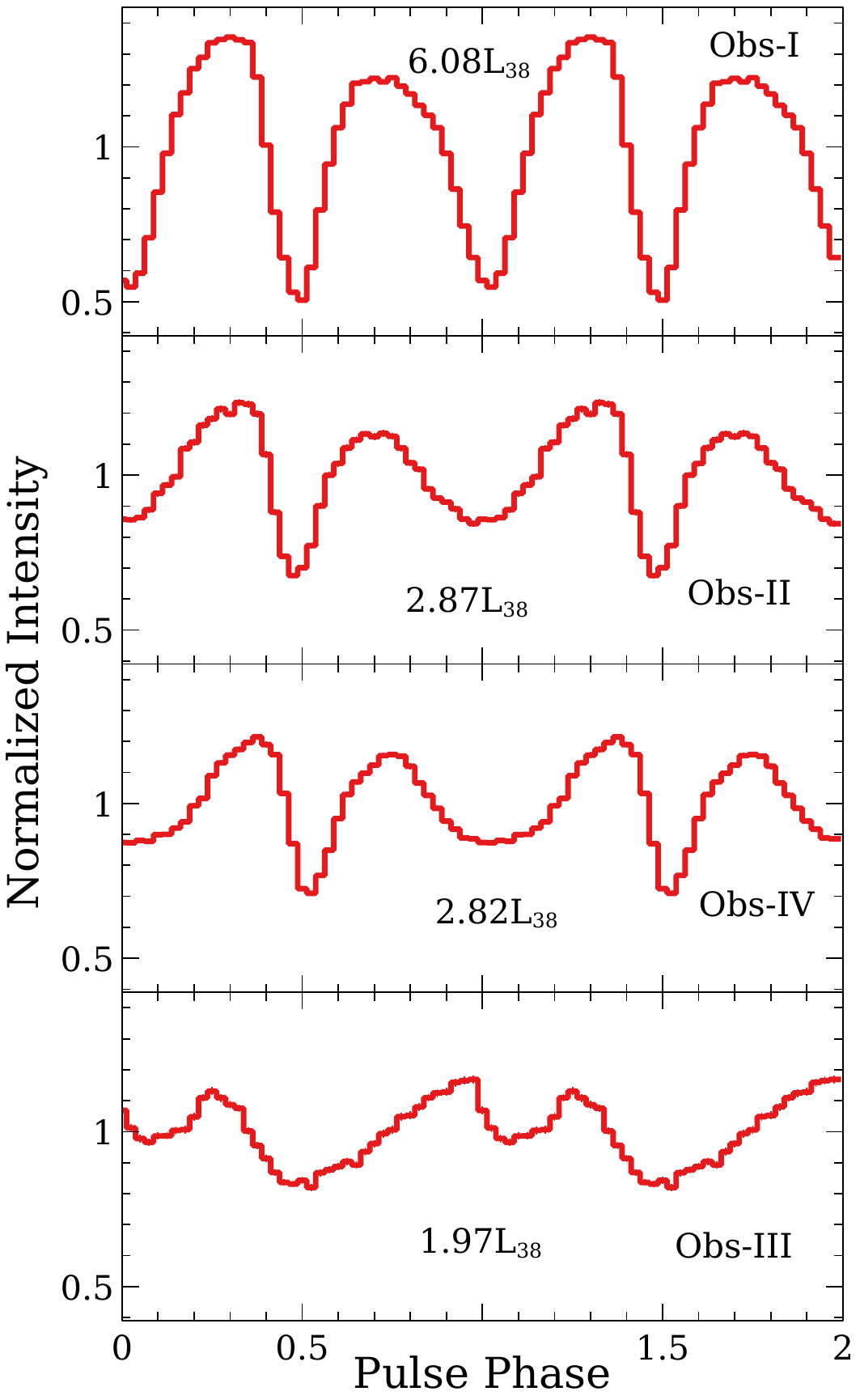}
\caption{The 3--79 keV pulse profiles of SMC~X-2, observed with {\it NuSTAR}, arranged in order of luminosity. The top two (Obs-I and Obs-II) and bottom (Obs-III) panels are from the 2015 outburst. The third panel (Obs-IV) represents the pulse profile from 
the recent 2022 outburst. L$\rm_{38}$ stands for 0.5-100 keV unabsorbed luminosity in 10$^{38}$~ergs~s$^{-1}$ unit.}
\label{broad_pp}
\end{figure}
\begin{figure*}
\centering
	{\includegraphics[width=4.5cm, height=8cm]{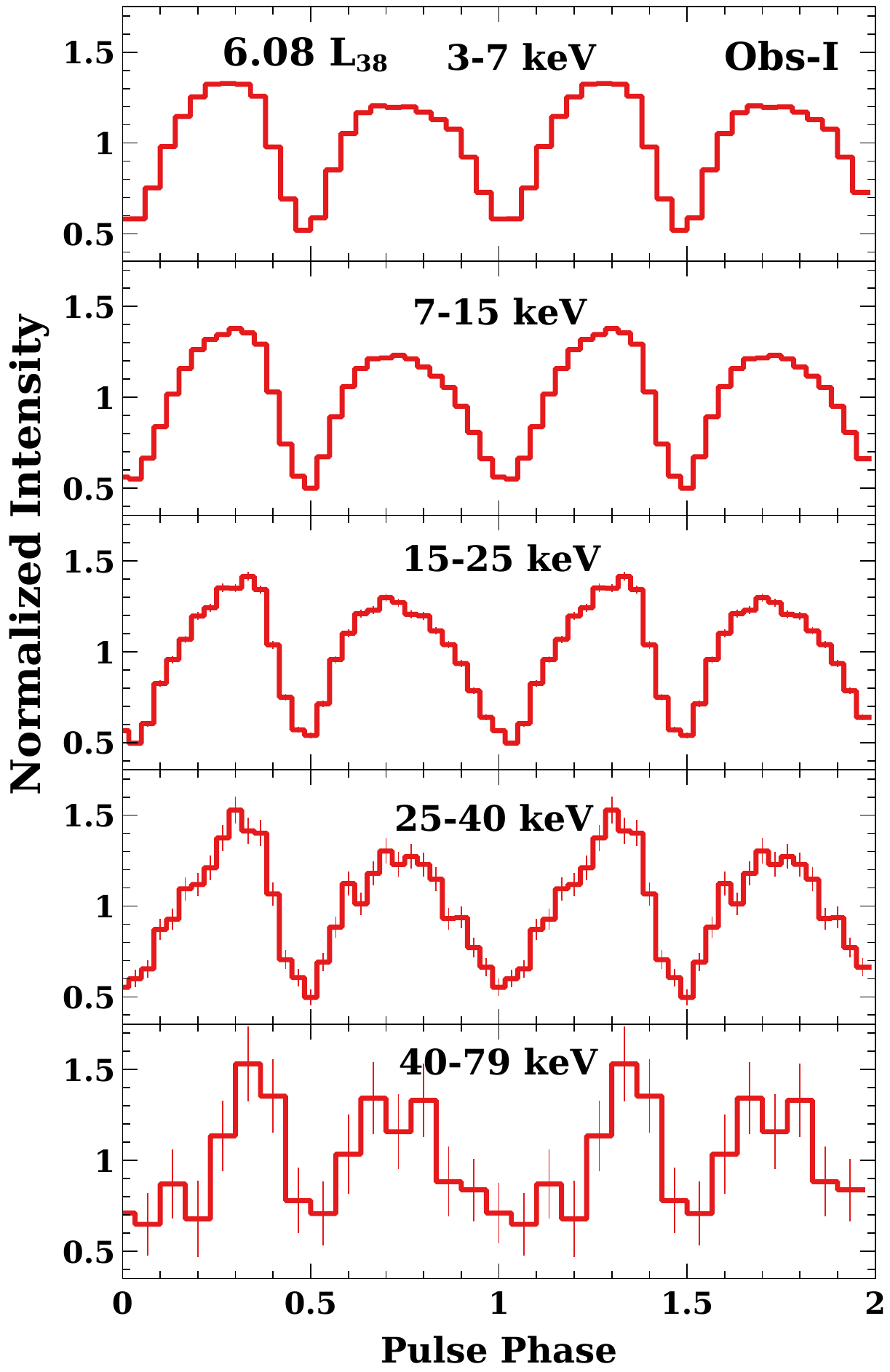}}
 {\includegraphics[width=4.2cm, height=8cm]{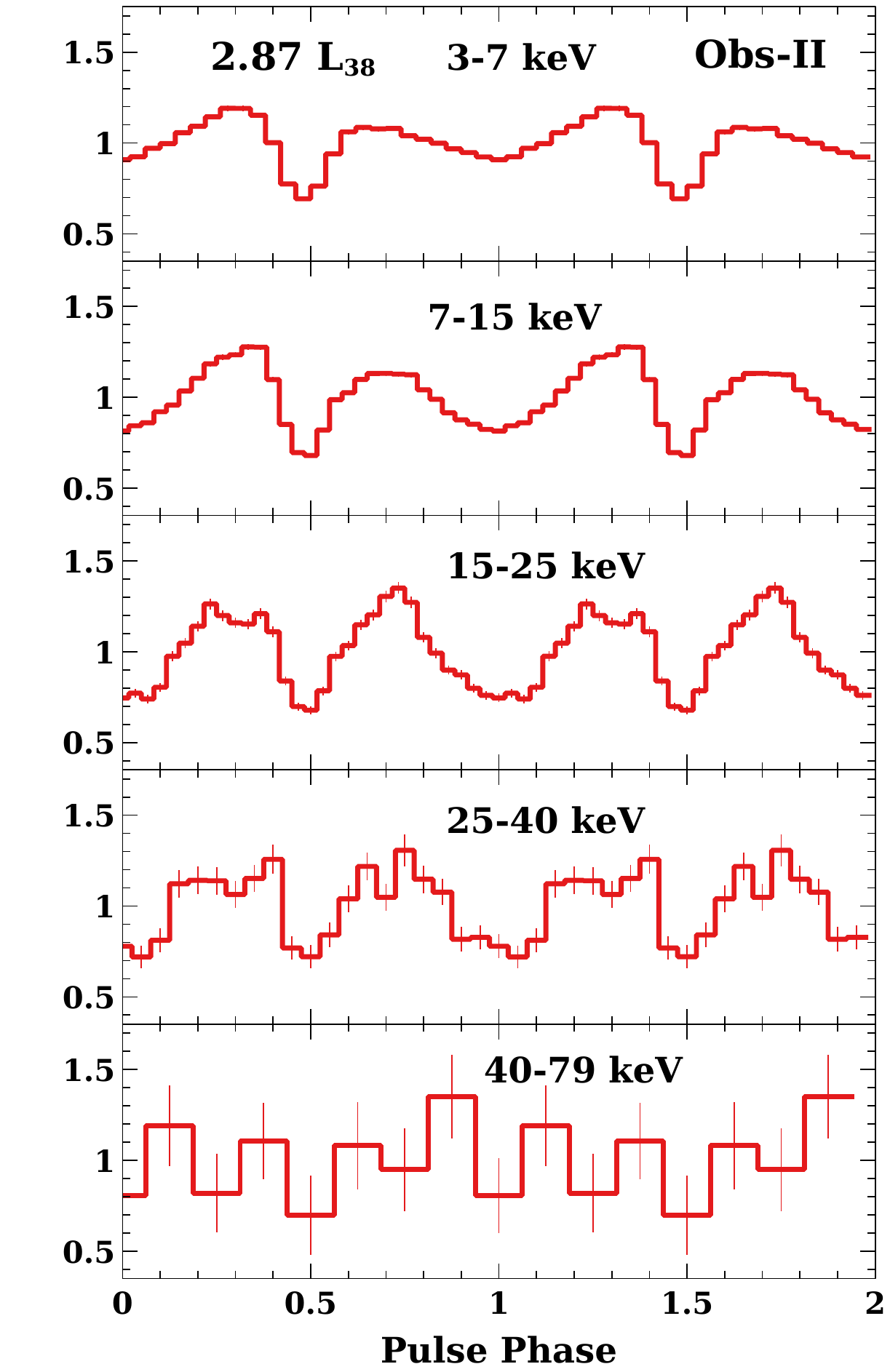}} 
		{\includegraphics[width=4.2cm, height=8cm]{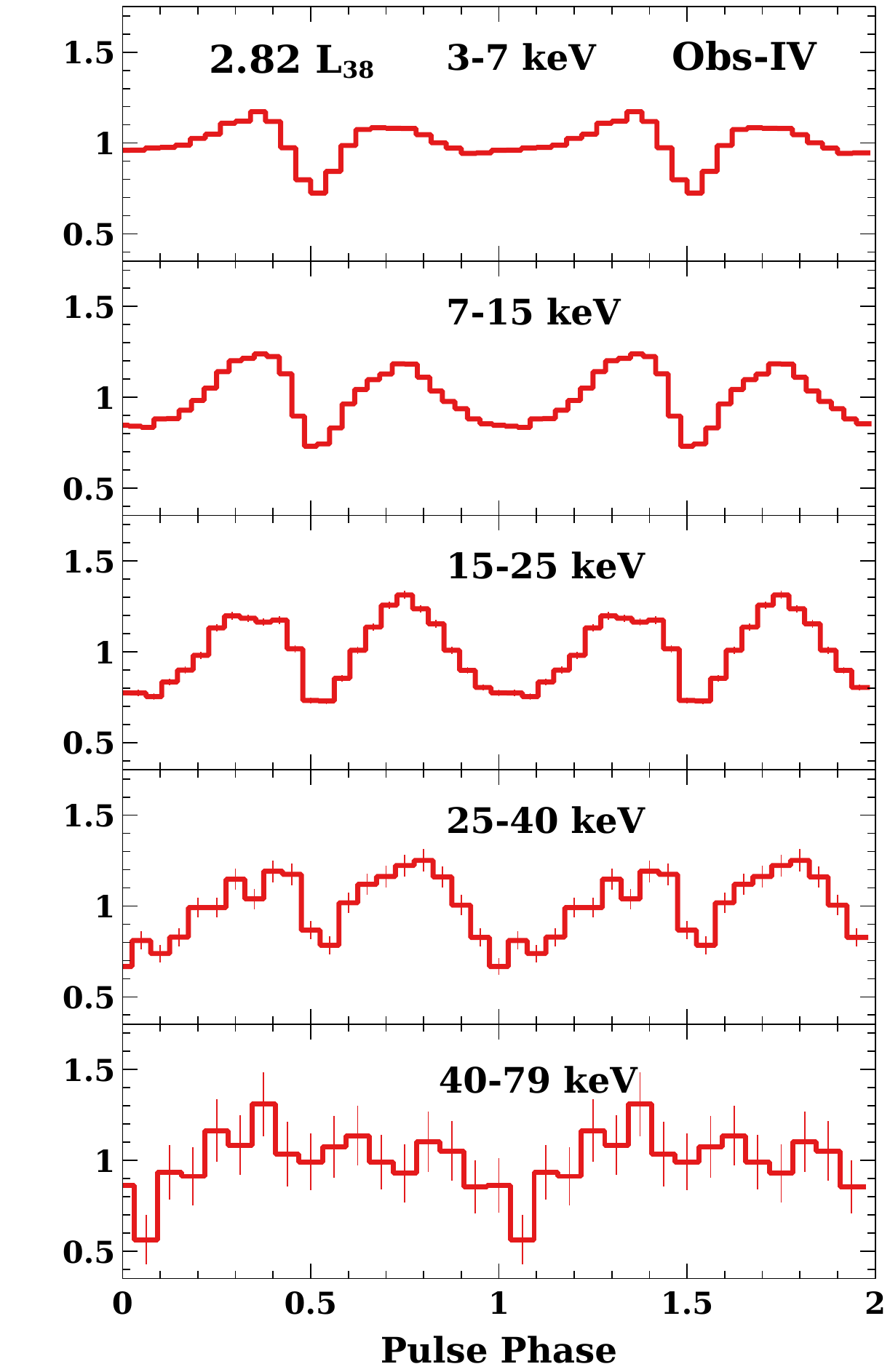}}		
  {\includegraphics[width=4.2cm, height=8cm]{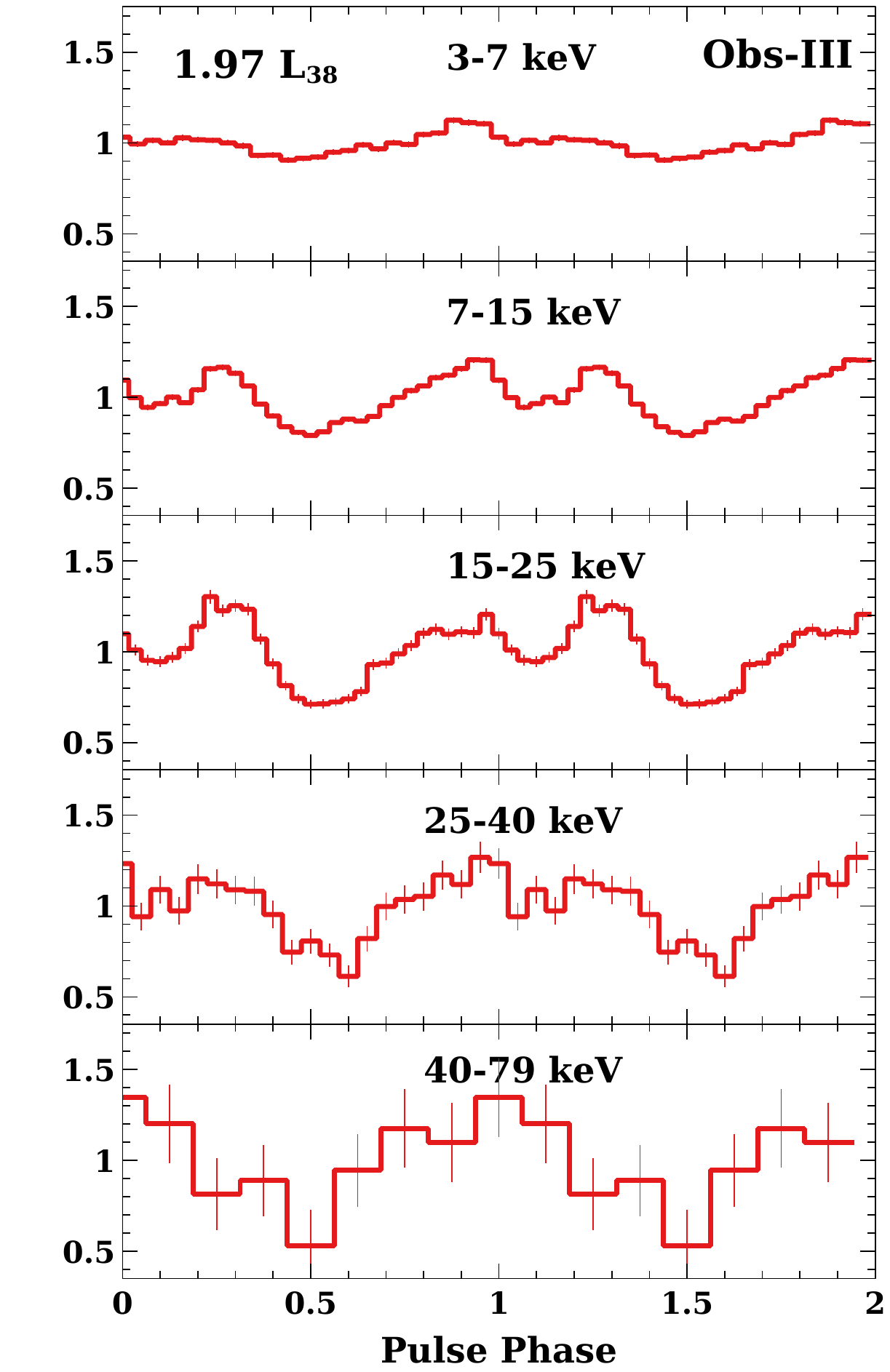}}
    \caption{The energy resolved pulse profiles of SMC~X-2 from the \nustar observations during its 2015 (Obs-I, Obs-II, and Obs-III) and 2022 (Obs-IV) outbursts. L$\rm_{38}$ stands for 0.5-100 keV unabsorbed luminosity in units of 10$^{38}$~ergs~s$^{-1}$.}
    \label{fig:erpp}
\end{figure*}

\section{Results}

\subsection{Timing analysis}

We searched for X-ray pulsations in the 3--79 keV \nustar light curves of \source using the $\chi^2$-maximization technique \citep{Leahy1987}. The barycentric corrected pulse period of the neutron star was estimated to be  2.37197(2), 2.37141(2),  2.37257(2), and  2.37283(1)~s during Obs-I, Obs-II, Obs-III, and Obs-IV, respectively. The light curves from all the \nustar observations are folded with the corresponding pulse period to obtain pulse profiles in the 3--79 keV band,  which are shown in  Figure~\ref{broad_pp}. A double-peaked profile is clearly evident at a luminosity\footnote{The 0.5--100 keV unabsorbed luminosity is calculated at a distance of 62 kpc after broadband spectroscopy that is presented in Section~3.2.} of 6$\times$10$^{38}$~ergs~s$^{-1}$ (top panel of Figure~\ref{broad_pp}). The shape of the pulse profile changes as luminosity decreases to $<$3$\times$10$^{38}$~ergs~s$^{-1}$ where a hump-like structure appears in the profile between 0.5--1.5 pulse phase. The pulse profiles exhibit similar morphology at an equal level of luminosities observed during the 2015 (Obs-II) and 2022 (Obs-IV) outbursts.

We adopted the following definition to evaluate the pulse fraction (PF) of the pulsar, i.e., the relative amplitude of pulsed modulation:
\begin{equation} \label{eq1}
PF = \frac{F_{max} - F_{min}}{F_{max} + F_{min}},
\end{equation}
where, {\em F$_{max}$} and {\em F$_{min}$} are the maximum and minimum intensities observed in the pulse profile, respectively. The estimated PFs during Obs-I, Obs-II, Obs-III, and Obs-IV are $\approx$45.7$\pm$0.4, 29.2$\pm$0.4, 17.6$\pm$0.2, and 26.2$\pm$0.3\%, respectively. The observed PF is correlated with luminosity. This means that the pulsed emission from the accretion column or the hot spot contributes more than the unpulsed radiation (originating from accretion flow or from the surface) when the source luminosity increases.

\begin{figure}
\centering
\includegraphics[height=3.3in, width=3.2in]{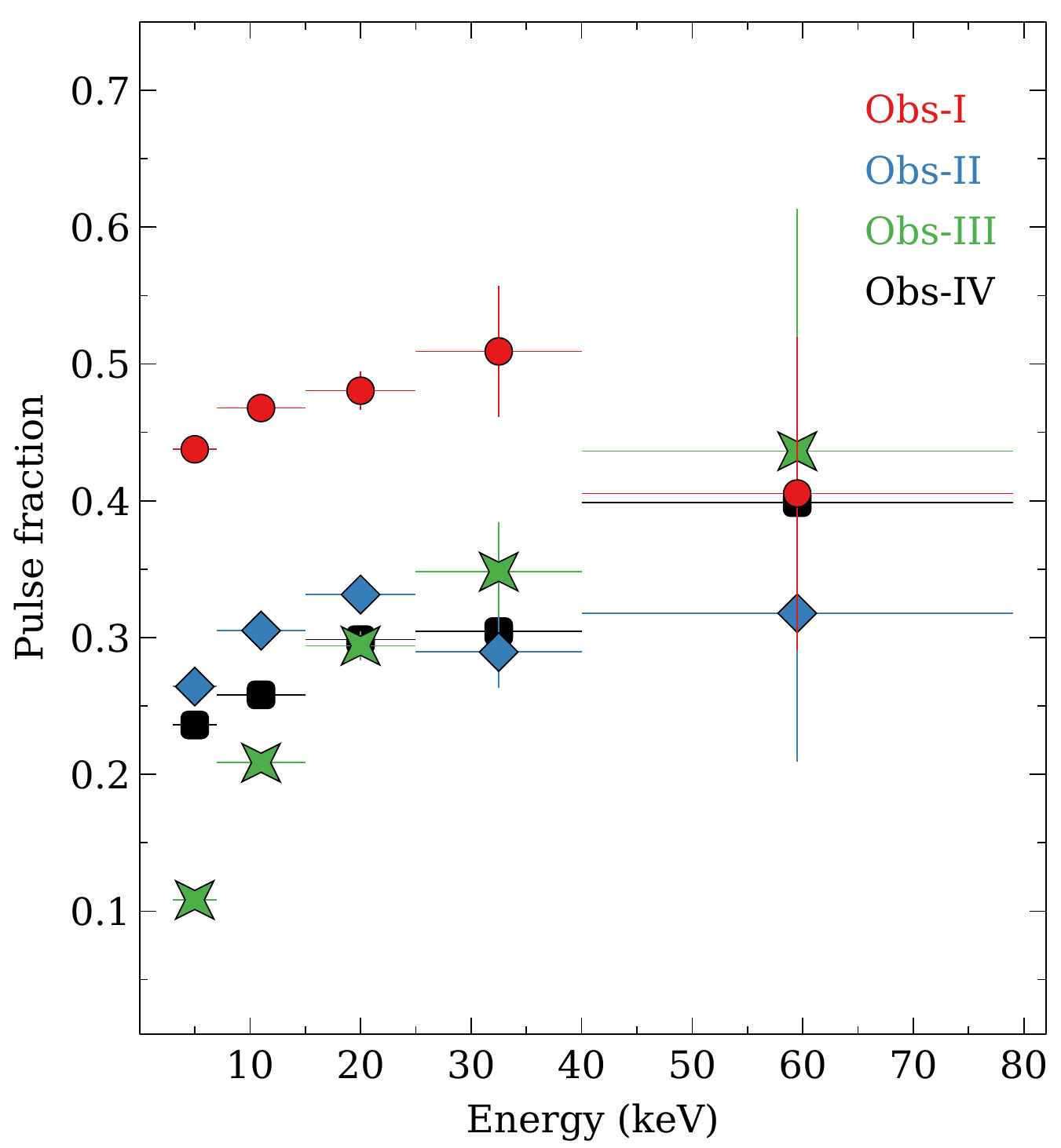}
\caption{The evolution of pulse fraction with energy, obtained from the energy resolved pulse profiles of the pulsar.}
\label{fig:pf}
\end{figure}

To examine the evolution of pulsed beam geometry of the source, we folded the energy-resolved light curves in 3--7, 7--15, 15--25, 25--40, and 40--79 keV from the \nustar observations (Figure~\ref{fig:erpp}). Doubled peaked profiles are observed throughout soft to hard X-ray energy ranges during Obs-I. The emission from both poles is clearly apparent at this stage. As the luminosity decreases to a level of $\sim$3$\times$10$^{38}$~ergs~s$^{-1}$, the pulse profiles below 15 keV (Obs-II and Obs-IV) appear with a hump-like structure that further evolves into a double-peak at higher energies. The Obs-III is at around the Eddington limit of a classical neutron star. At this point, the soft to hard pulse profiles are single-peaked with a hump in the middle. We further studied the energy evolution of the PF from the pulse profiles. Figure~\ref{fig:pf} shows positive dependencies between the energy and the PF. An extremely low PF of about 11\% is found during Obs-III in the 3--7 keV band.

\begin{figure}
\centering
\includegraphics[width=\columnwidth]{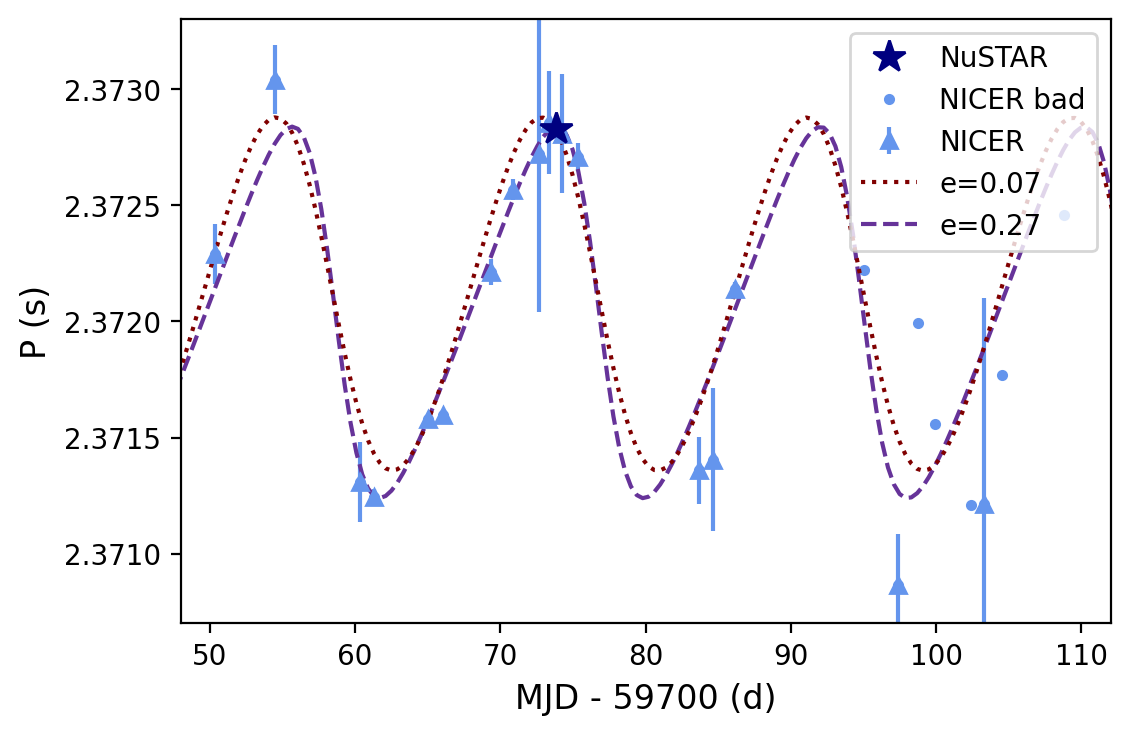}
\includegraphics[width=\columnwidth]{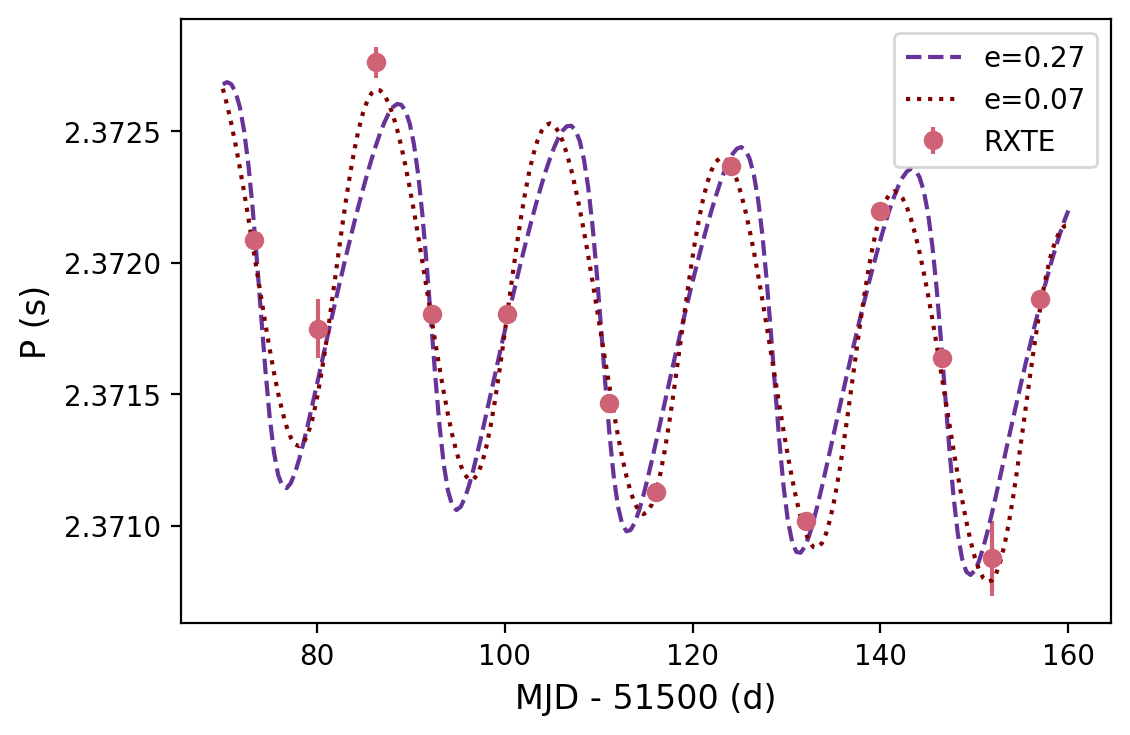}
\caption{Period evolution of SMC X-2 during the 2022 outburst using \nicer~ and \nustar observations in top panel. The bottom panel shows the pulse period evolution observed by \rxte  during the 2000 January-May outburst.}
\label{P_evol}
\end{figure}

\subsection{Spin evolution}

We used the barycentric corrected \nicer data to investigate the evolution of the temporal properties with the outburst and in particular, the spin evolution.
To search for a periodic signal, we followed the same methodology adopted for \nicer monitoring of SXP 15.6 \citep[see][]{2022A&A...664A.194V}.
First we adopted the epoch folding Z-search test implemented through {\tt HENdrics} and {\tt Stingray} \citep[][]{2019ApJ...881...39H}.
Then we refined the temporal solution and its uncertainty based on the ToA of individual pulses by using PINT\footnote{\url{https://github.com/nanograv/pint/}} \citep{2021ApJ...911...45L}.
For some snapshot phases connecting the ToAs was challenging to impossible due to multiple peaks in the periodogram with similar intensity, a problem often encountered in slow pulsars observed with gaps \citep[e.g. see appendix of][]{2017ApJ...839..125Z,2018A&A...620L..12V}.
The resulting period measurements are shown in Fig. \ref{P_evol}, In the same plot, we mark the \nustar measurement. We note that \nicer observations where a defined period was not obtained are marked as `bad` and no error is given as their uncertainty is similar to the range of the plot. 

In the spin evolution, we see clear signatures of orbital motion, consistent with the 18.38 d period reported in the literature based on \rxte data \citep{Townsend2011MNRAS.416.1556T}, while a similar periodicity is also seen in optical data of the system \citep[e.g.][]{Roy2022ApJ...936...90R}.
Modeling of the archival \rxte data has revealed an almost circular orbit with an eccentricity of about 0.07. The sampling of the \rxte data was sparse and in similar orbital phases. In contrast, the new \nicer data cover a smaller baseline but better sample the orbital cycle near the peak of the outburst. Thus their study could help improve the old orbital solution.
To model the orbital evolution, we used a model composed of the orbital signature and the first derivative of the frequency similar to \citet{Townsend2011MNRAS.416.1556T}. 
Moreover, we followed a Bayesian approach to effectively sample the  whole parameter space of the underline model \citep{Karaferias2023MNRAS.520..281K}. 
We applied the same method to the \nicer data, and the \rxte measurements\footnote{Data points were extracted from \citet{Townsend2011MNRAS.416.1556T}}.
Given the issues with the \nicer period determination at the later stages of the 2022 monitoring, we constrained our fit  within MJD 59750-59790.
We find that the 2022 monitoring data yield a somehow eccentric orbit (e=0.27), while the period is better constrained by the  \rxte data. Interestingly, if we maintain no priors for the eccentricity the fit to the  \rxte data yields a lower eccentricity consistent with the one reported by \citet{Townsend2011MNRAS.416.1556T}. For comparison, we plotted both solutions to the newly obtained and archival data shown in Figure \ref{P_evol}.

\begin{figure}
\centering
\includegraphics[height=2.8in, width=3.25in]{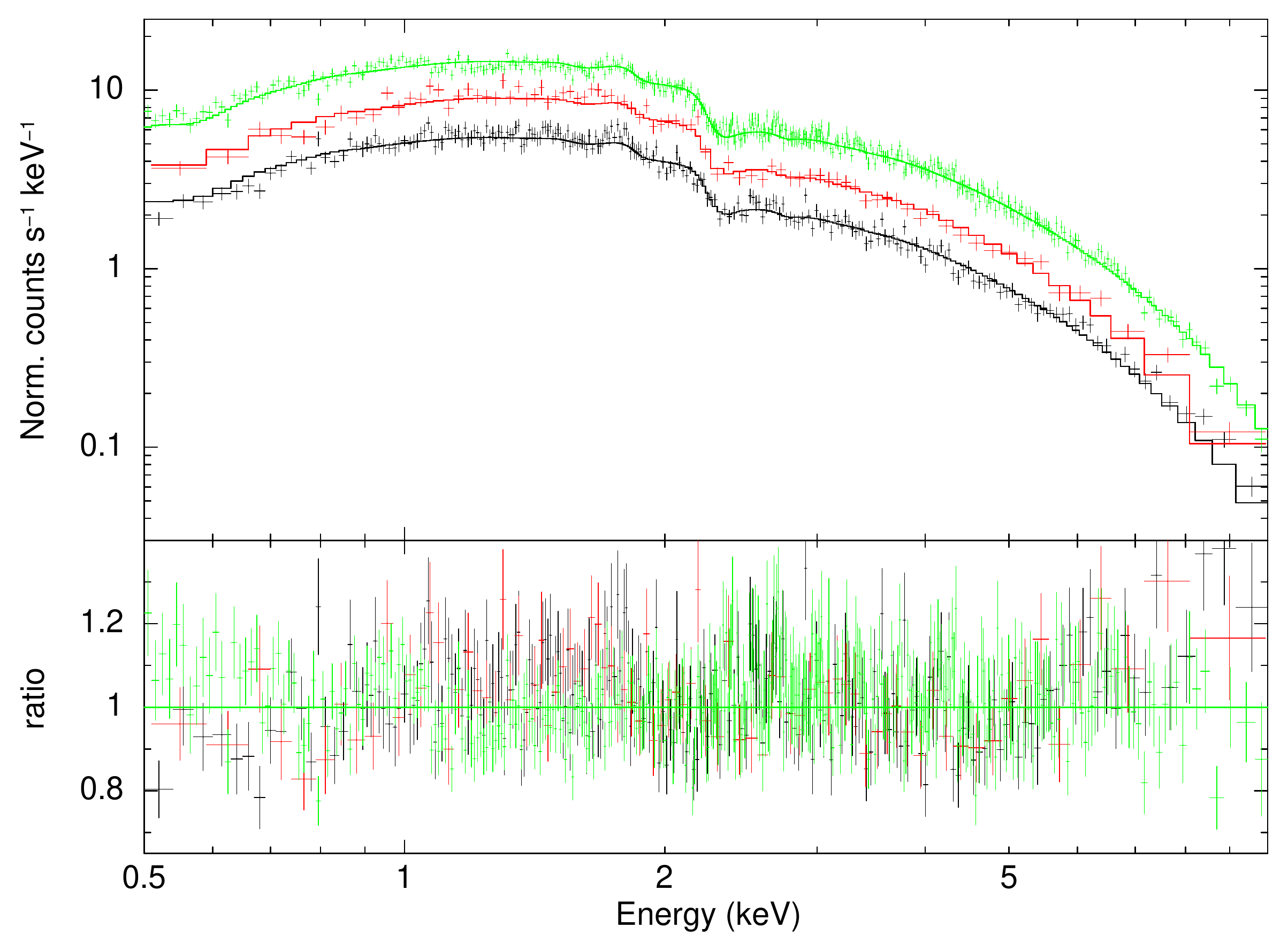}
\caption{Three representative \nicer spectra in 0.5--10 keV fitted with an absorbed cutoff power-law model. Black, red, and green correspond to an unabsorbed luminosity of  4.4$\times$10$^{37}$, 6.8$\times$10$^{37}$, and 1.3$\times$10$^{38}$~ergs~s$^{-1}$ in the 0.5--10 keV range, respectively.  }
\label{fig:nicer_spec}
\end{figure}
\begin{figure}
\centering
\includegraphics[height=4.7in, width=3.2in]{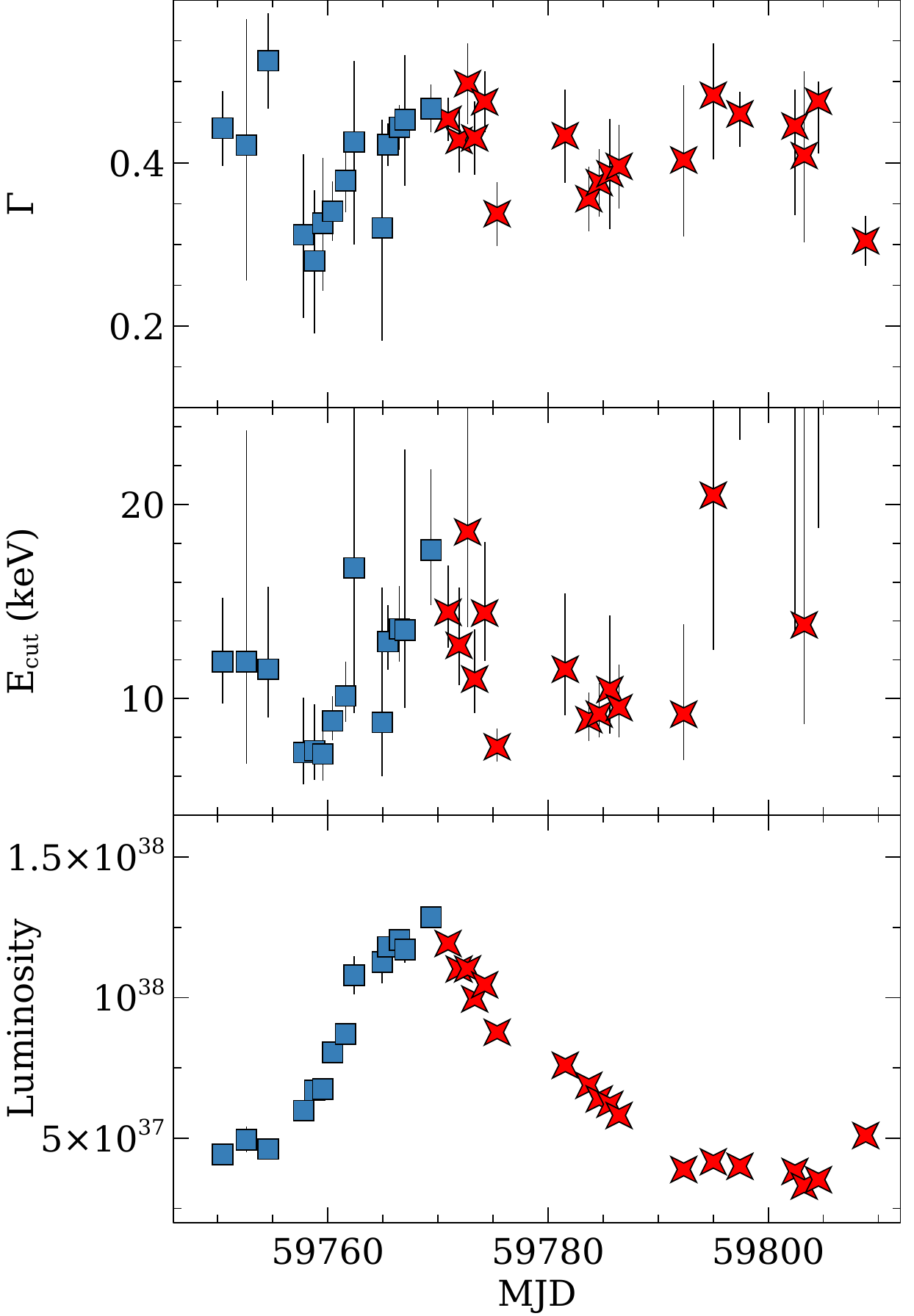}
\caption{Spectral parameters of SMC~X-2 after fitting each \nicer energy spectrum with an absorbed cutoff power-law model. The absorption column density is fixed at 1.43$\times$10$^{21}$ cm$^{-2}$. The unabsorbed luminosity is estimated in 0.5-10 keV range for a source distance of 62 kpc. The outburst peaked around 2022 July 10 with a luminosity of 1.3$\times$10$^{38}$~ergs~s$^{-1}$ in the \nicer band. The square (blue) and squashbox (red) symbols correspond to the rising and declining parts of the 2022 outburst, respectively.}
\label{fig:spec-para}
\end{figure}

\begin{figure}
\centering
\includegraphics[height=4.7in, width=3.2in]{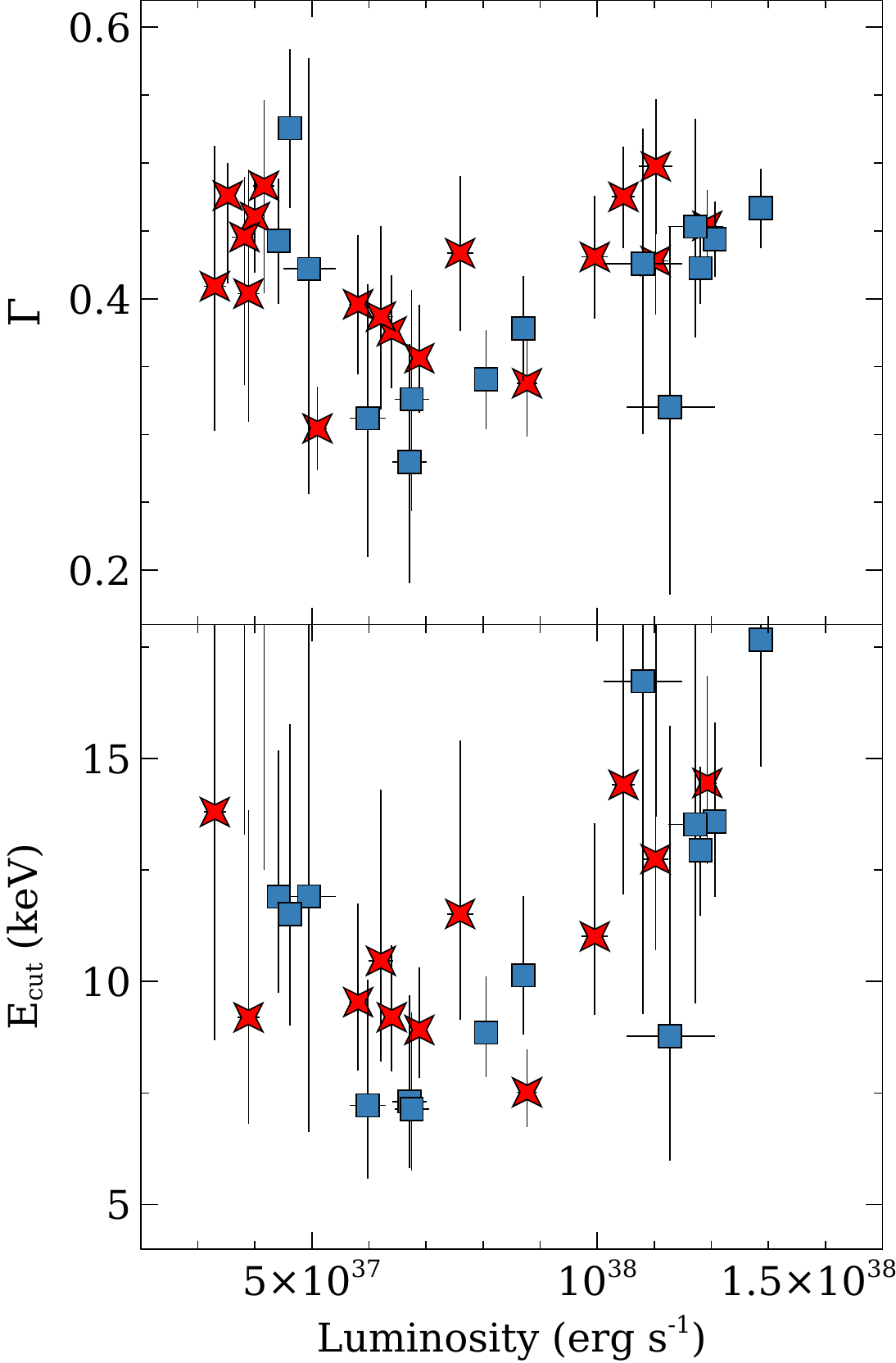}
\caption{The evolution of photon index and cutoff energy with the 0.5-10 keV luminosity of the pulsar. Square (blue) and squashbox (red) symbols correspond to the rising and declining parts of the 2022 outburst, respectively }
\label{fig:spec-para2}
\end{figure}

\begin{figure*}
\centering
	{\includegraphics[width=8.5cm, height=8.cm]{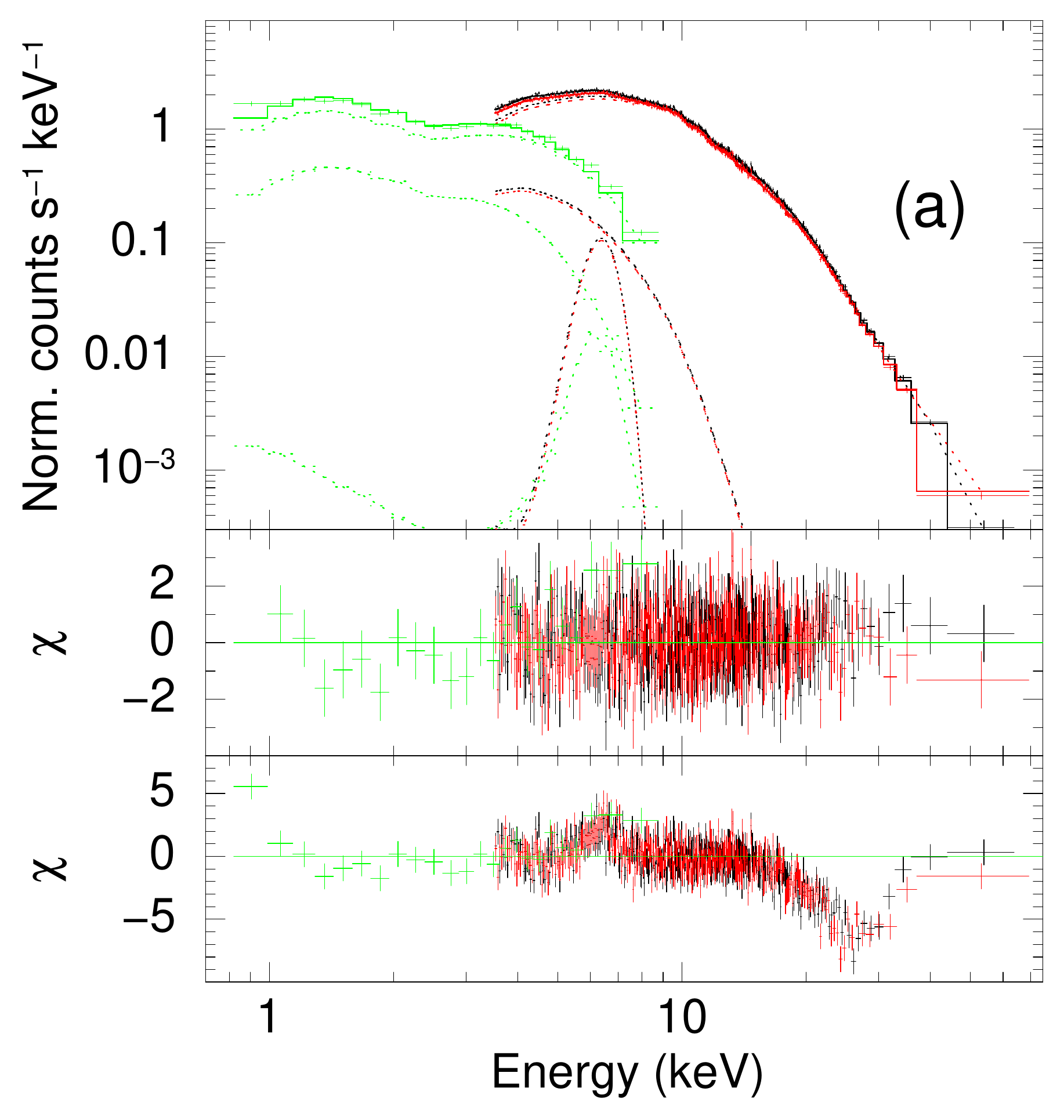}}
 {\includegraphics[width=8.5cm, height=8.cm]{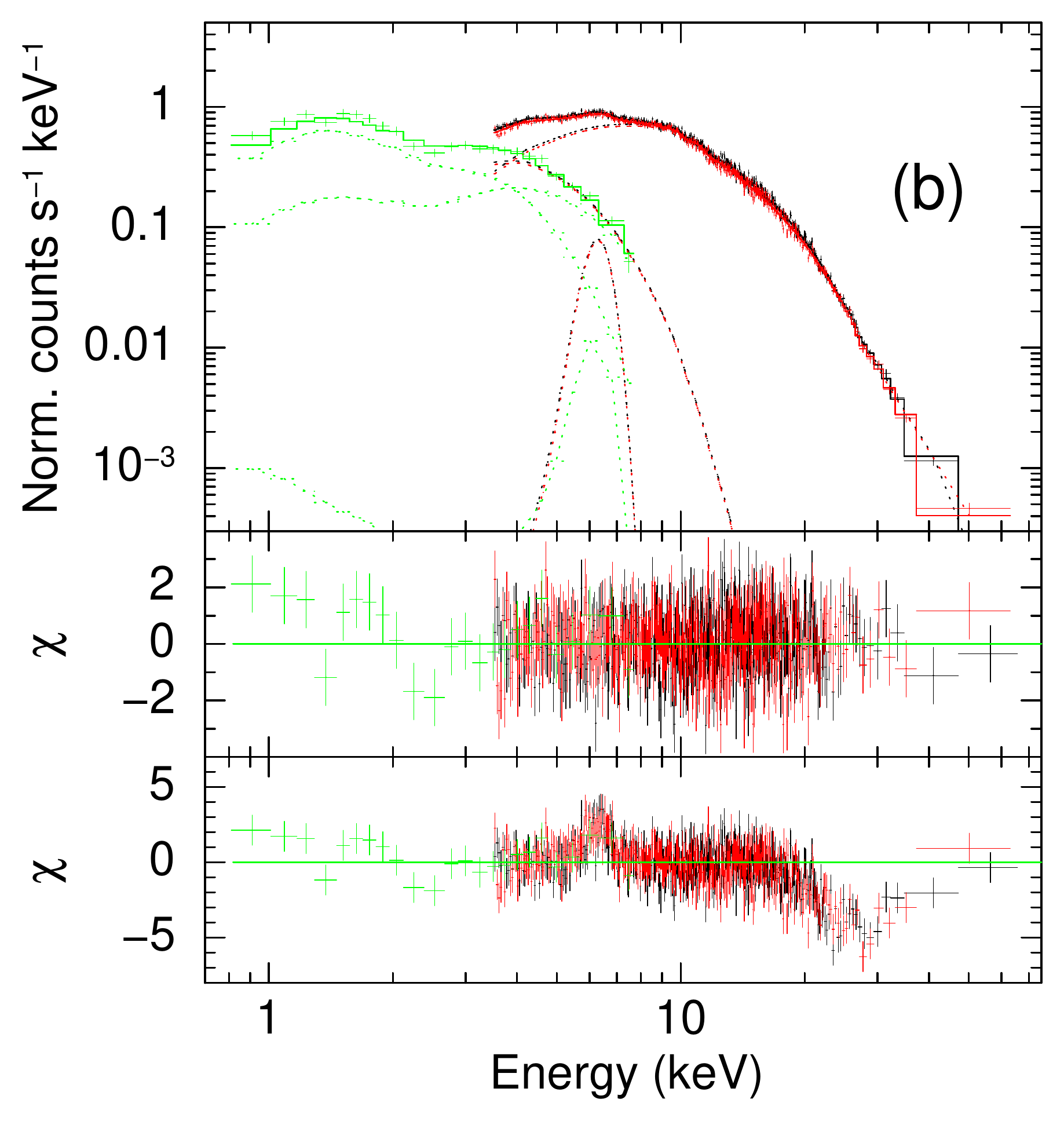}} \\
		{\includegraphics[width=8.5cm, height=8.cm]{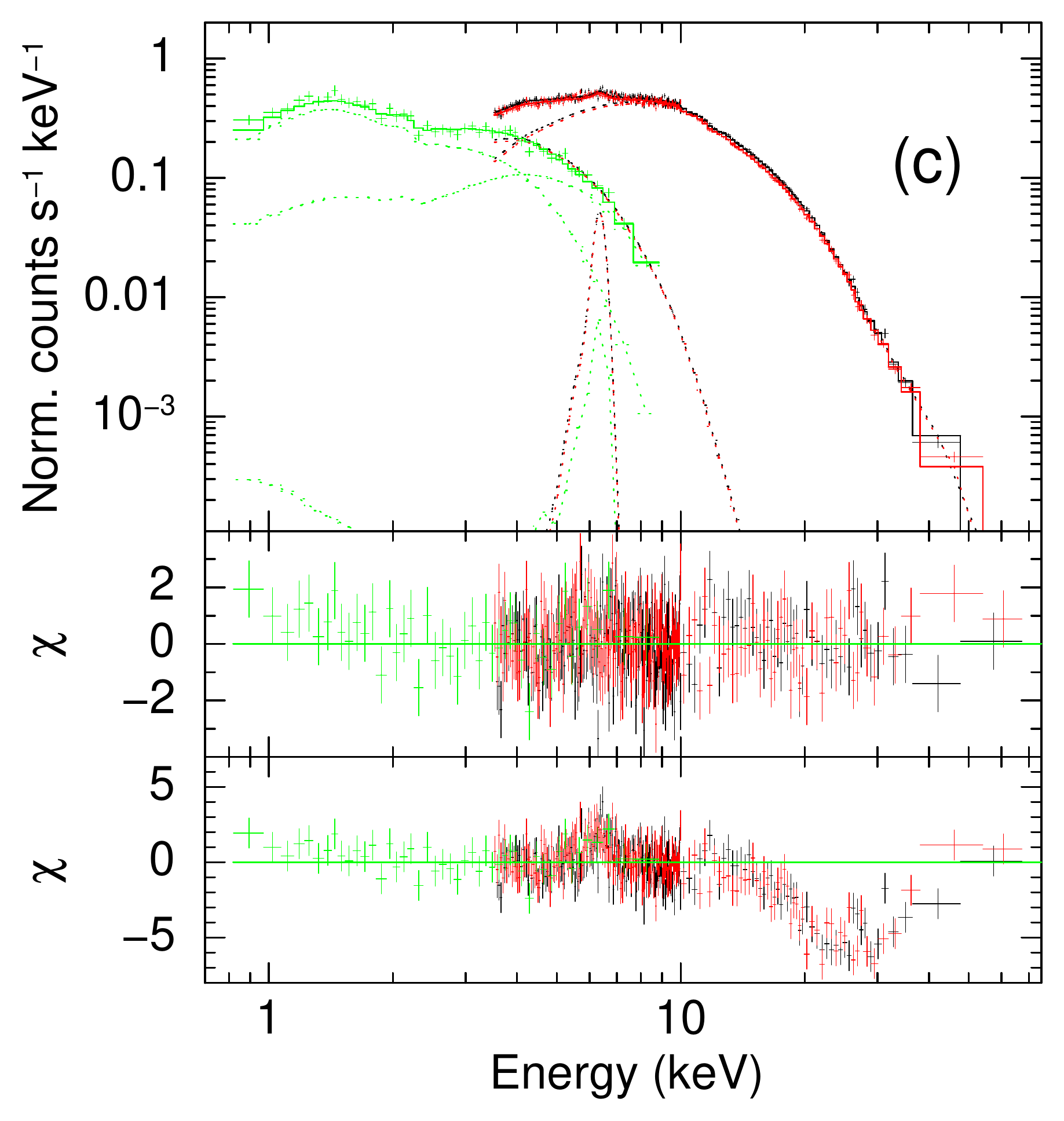}}
		{\includegraphics[width=8.5cm, height=8.cm]{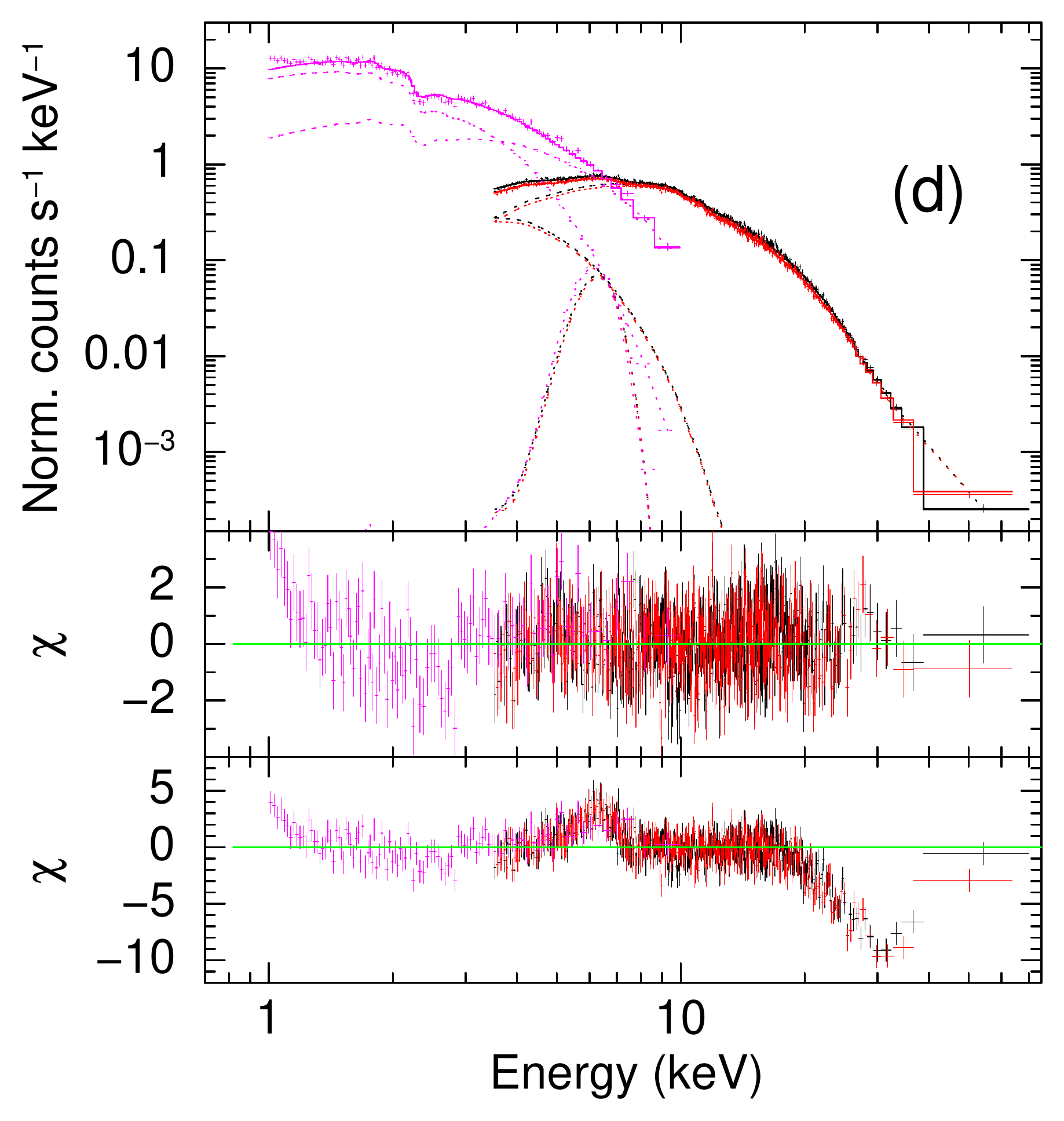}}
    \caption{The broadband energy spectra of the pulsar during its 2015 and 2022 outbursts are fitted with  an absorbed cutoff power-law model along with a blackbody, and an iron emission line as well as a Gaussian absorption feature for a cyclotron line (top).  The middle and bottom panels of each figure correspond to the spectral residuals with and without an iron line and cyclotron feature component to the best-fitted spectra, respectively. (a), (b), (c), and (d) denote the sets of \nustar and Swift/XRT (green) or \nicer (magenta) data in Table~1, respectively. }
    \label{fig:joint_spec}
\end{figure*}

\subsection{The continuum emission during 2022 outburst}

The soft X-ray energy spectrum of \source is studied to understand the emission at different phases of the 2022 outburst. We used \nicer observations at multiple epochs of the outburst in this study. Each 0.5--10 keV \nicer spectrum can be fitted statistically well with an absorbed cutoff power-law model in XSPEC (\citealt{Arnaud1996}). Figure~\ref{fig:nicer_spec} shows the energy spectra from these observations together. We considered {\tt TBabs} model at  Wilm abundance \citep{Wilms2000} and Verner cross-section \citep{Verner1996ApJ...465..487V} to account for the local and interstellar medium absorptions along the source direction in our spectral fit. We found that the equivalent hydrogen column density (N$_{\rm H}$) varies  in the range of (1--3)$\times$10$^{21}$ atoms cm$^{-2}$ during these observations. To reduce the spectral fitting degeneracy in the soft X-ray band, the column density is fixed at an average value of 1.43$\times$10$^{21}$ cm$^{-2}$, which is consistent with the value observed with XMM-Newton in 2015 within error bars \citep{Palombara2016MNRAS.458L..74L}. We then fitted each individual spectrum with an absorbed cutoff power-law model at the above-fixed column density. No signature of iron fluorescence line is detected in the \nicer spectra. The 0.5--10 keV unabsorbed flux is estimated using the {\tt cflux} convolution model. The parameter errors are estimated for the 90\% confidence interval in this paper.  

The evolution of the spectral parameters during the outburst is shown in Figure~\ref{fig:spec-para}. The power-law photon index and cutoff energy are variable across the rise (blue) and decay (red) phases of the outburst. The peak luminosity in the 0.5--10 keV range is found to be 1.3$\times$10$^{38}$~ergs~s$^{-1}$ (for a source distance of 62~kpc) around the outburst peak. To examine the spectral changes and associated spectral transition, the parameters are studied with respect to the 0.5--10 keV unabsorbed luminosity in Figure~\ref{fig:spec-para2}. The photon index and cutoff energy follow a positive correlation   (with a Pearson correlation coefficient of 0.61 and 0.66, respectively) with luminosity above 9$\times$10$^{37}$~ergs~s$^{-1}$. The index stays close to a value of 0.35 at luminosity in the range of (6-9)$\times$10$^{37}$~ergs~s$^{-1}$. This could be identified as a critical limit of the pulsar where the spectral shape changes. Below this limit, the source shows a negative correlation (with a Pearson correlation coefficient of --0.57) between luminosity and the photon index.  Similar behavior is observed between luminosity and the cutoff energy (with a Pearson correlation coefficient of --0.54).  

 We also fitted each \nicer spectrum with two absorption components together with a cutoff power-law model. One of the column densities was fixed at a foreground value 4$\times$10$^{20}$~cm$^{-2}$ \citep{Dickey1990ARA&A..28..215D} at the Solar abundance. Another column density was free in our study at the SMC abundance of 0.2 Solar \citep{Russell1992ApJ...384..508R} to incorporate the local absorption. The SMC column density changes in the range (0.5--3)$\times$10$^{21}$~cm$^{-2}$ during the outburst.  All the fitted parameters, such as SMC column density, photon index, and the cutoff energy, however, were found to decrease with luminosity.  A gradual decline may be because of spectral degeneracy among these parameters in the soft X-ray band. This is similar to the findings discussed above in this section. Though we can not totally disentangle the variation of the local absorption during the outburst, therefore, to reduce the degeneracy among the parameters, we fixed the SMC column density to an average value of 1.25$\times$10$^{21}$ cm$^{-2}$. This resulted in similar kinds of parameter variations as reported in Figures~\ref{fig:spec-para} \& \ref{fig:spec-para2}.

\subsection{Broadband spectroscopy of the pulsar during 2015 and 2022 giant outbursts}

\begin{figure}
\centering
	{\includegraphics[width=8.2cm, height=6.7cm]{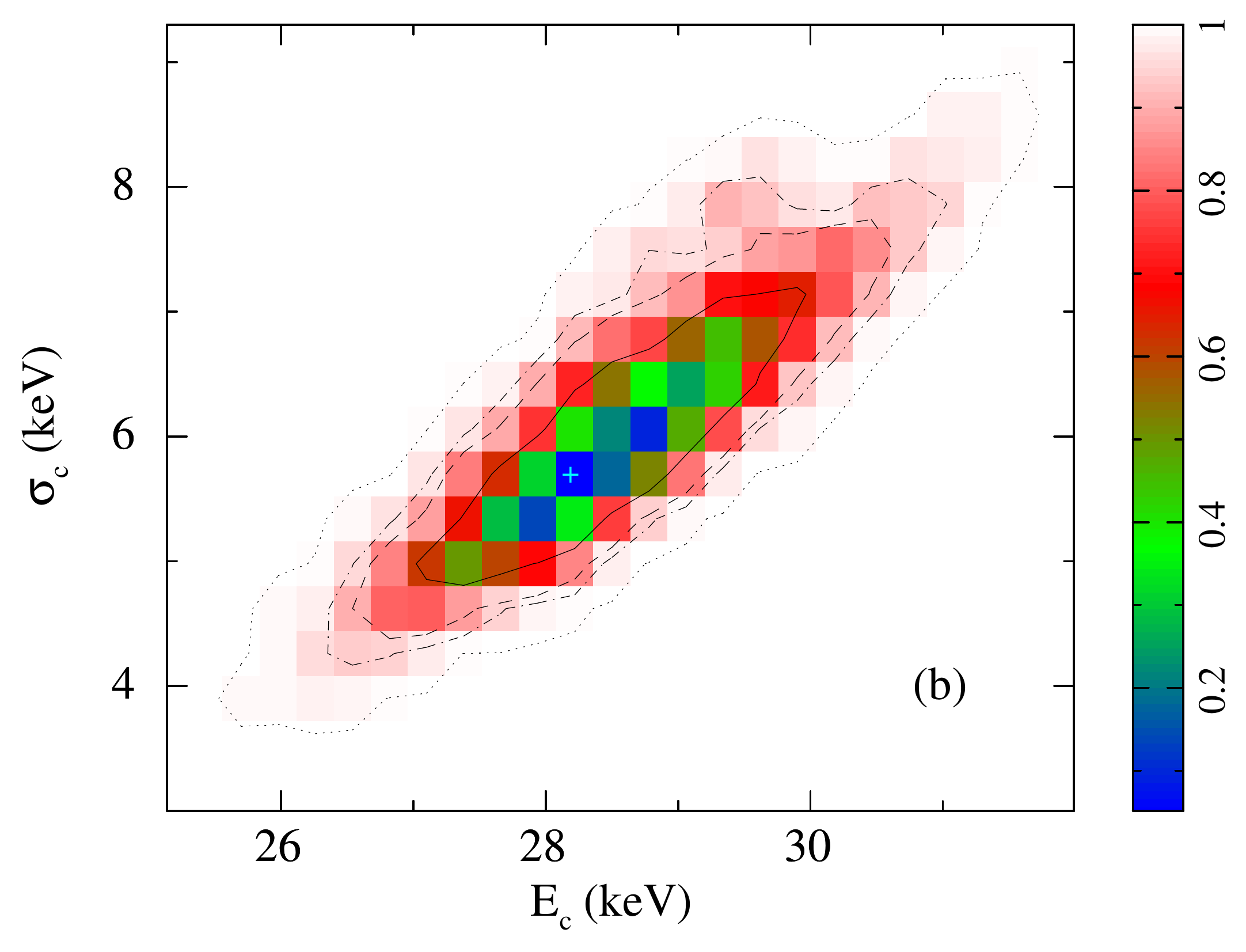}}
 {\includegraphics[width=8.2cm, height=6.7cm]{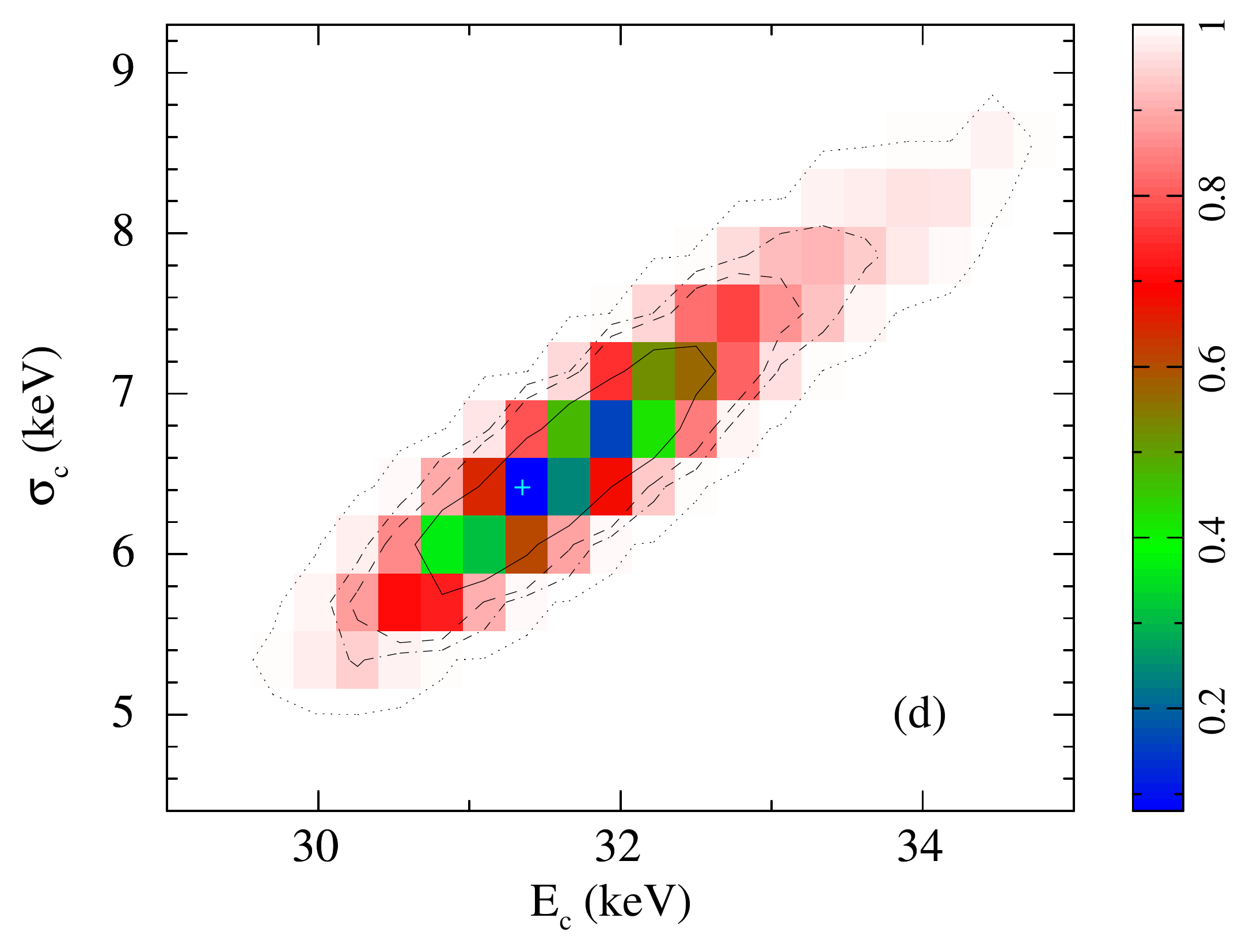}} 
    \caption{The 68, 90, 95, and 99.7\% (from innermost to outermost) confidence intervals contour maps between cyclotron line energy and width at an equivalent level of luminosity from 2015 (upper) and 2022 (bottom) outbursts. (b) and (d) have the usual meaning as denoted in Figure~\ref{fig:joint_spec}. The `+' sign corresponds to the best-fitting values for both parameters. The color bars denote the parameter confidence range from 68 to 99.7\% on a scale of zero to one, respectively.}
    \label{fig:cntmap}
\end{figure}

We further performed broadband spectroscopy of \source in the 1--70 keV range to investigate the continuum as well as the nature of the cyclotron absorption line and its evolution. Three sets of XRT and \nustar quasi-simultaneous data from the 2015 outburst are used along with the 2022 July \nicer and \nustar pointed observations (Figure~\ref{fig:joint_spec}). These spectra are additionally binned for representation purposes using {\tt setplot rebin} command of {\tt XSPEC}. The pulsar spectrum can be described by standard models such as the Fermi-Dirac cutoff power-law, the Negative and Positive power-law with Exponential cutoff (NPEX; \citealt{Makishima1999ApJ...525..978M}), and the cutoff power-law with a blackbody component (see also \citealt{Jaisawal2016MNRAS.461L..97J}). In our study, we used an absorbed cutoff power-law model with a blackbody component ({\tt bbodyrad} in {\tt XSPEC})  to study all four data sets.  The column density was pegged at a lower limit of 1$\times$10$^{20}$~cm$^{-2}$ in the fit. A 6.4 keV Gaussian component for an iron fluorescence line is also needed. In addition to the emission, a cyclotron absorption feature appears in the data and is fitted with a Gaussian absorption line. The cyclotron lines are detected at $<$30 keV and at 31.5 keV in the 2015 and 2022 data sets, respectively (Table~\ref{spec_par}).  We used the Markov chain Monte Carlo (MCMC) model in {\tt XSPEC} for error estimation on the spectral parameters. The Goodman–Weare algorithm with 20 walkers and a total length of 200000 is considered in our study. The spectral parameters at a 90\% confidence interval are presented in Table~\ref{spec_par}. We also show the 68, 90, 95, and 99.7\% confidence level contour maps for cyclotron line energy and width for second and fourth \nustar observations in Figure~\ref{fig:cntmap}.

\begin{figure}
\centering
\includegraphics[height=4.25in, width=3.2in]{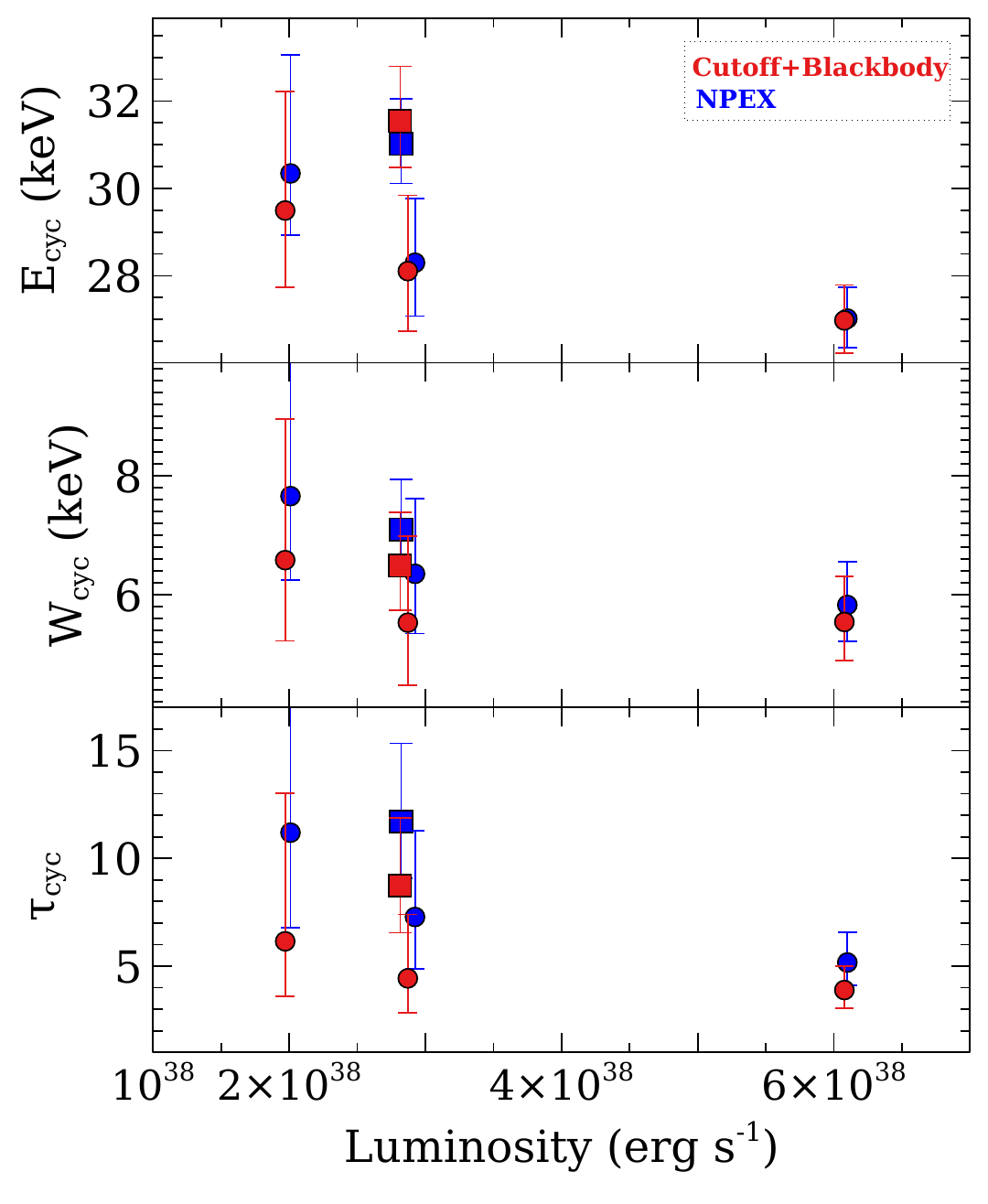}
\caption{Change in the cyclotron line parameters such as line energy, width, and its strength ($\tau_{cyc}$) with 0.5--100 keV unabsorbed luminosity. The measurements from the 2015 and 2022 outbursts are shown in solid circles and squares, respectively The red and blue colors correspond to line parameters obtained after an absorbed cutoff power-law model along with blackbody and GABS, and an absorbed NPEX model with a GABS model, respectively.}
\label{fig:cyclotron}
\end{figure}

The cyclotron line energy, width, and line strength decreased with the luminosity during the 2015 outburst (Figure~\ref{fig:cyclotron}). The second and fourth \nustar observations (see Table~\ref{spec_par}) are at almost the same luminosity level. However, the cyclotron line parameters, especially the cyclotron line energies, are found to be marginally higher (within 90\% level) in 2022 July. The magnetic field of the neutron star can be estimated to be 3.5$\times$10$^{12}$~G based on this recent detection. 
Moreover, to study the model-independency of the cyclotron line, we show the parameter evolution using an absorbed NPEX model with a GABS component. The cyclotron parameters are consistent between both models (Figure~\ref{fig:cyclotron}).     

Following the broadband fit, we found that the 0.5--10 keV luminosity represents about 40\% of the total pulsar emission in a 0.5--100 keV band. Therefore, an average bolometric correction of 2.5 can be applied to the outburst emission measured by \nicer. This is consistent with the typical estimates where the soft band mostly contributes around 30--40\% of the pulsar emission \citep{2022MNRAS.513.1400A,2022A&A...664A.194V}.
The 2022 outburst of \source peaked at around MJD 59769.4 with a bolometric corrected luminosity of 3.2$\times$10$^{38}$~ergs~s$^{-1}$. This is close to half of the peak value the pulsar attained in the 2015 outburst.  

 Other parameters in Table~\ref{spec_par} such as the blackbody temperature was found to be in a narrow range of (0.85--1.1) keV during both outbursts. We also measured the corresponding emission radius (Radius$_{\rm bb}$) to examine the possible origin and location of thermal emission. The observed Radius$_{\rm bb}$ is estimated to be in the range of (14--22)~km (see Table~\ref{spec_par}), which is close to the size of a neutron star. This suggests that the thermal component may originate from the neutron star surface or from the accretion column during the 2015 and 2022 outbursts.

\begin{table*}
\centering
\caption{Best-fitting spectral parameters (with 90\% errors) of \source obtained from four simultaneous {\it NuSTAR} and \swift/XRT or {\it NICER} observations. The fitted model consists of an absorbed cutoff power-law continuum with a blackbody and a Gaussian function for the iron emission line and cyclotron absorption line components.}

\begin{tabular}{lcccccc}
\hline
Parameters                 &Obs-I+XRT             &Obs-II+XRT   &Obs-III+XRT       &Obs-IV+\nicer  \\

\hline

N$_{\rm H}$$^a$              &0.01$^{+0.01}_{-0.01}$  &0.01$^{+0.08}_{-0.01}$   &0.01$^{+0.04}_{-0.01}$    &0.01$^{+0.6}_{-0.01}$ \\
Photon index              &-0.16$\pm$0.06         &-1$\pm$0.1   &-1.33$\pm$0.1   &-0.77$\pm$0.1\\
E$_{\rm cut}$ (keV)	          &6.4$\pm$0.2           &5.2$\pm$0.3     &5$\pm$0.3      &5.9$\pm$0.3    \\
\\
kT$_{\rm bb}$ (keV)                &1.1$\pm$0.1            &0.95$\pm$0.04      &0.95$\pm$0.04  &0.85$\pm$0.02\\
Norm$_{\rm bb}$                     &5.5$\pm$1.5           &9.1$\pm$1.1        &5.8$\pm$0.7    &12.4$\pm$0.7\\
Radius$_{\rm bb}$ (km)              &14.5$\pm$2              &18.7$\pm$1.1     &14.9$\pm$0.9   &21.8$\pm$0.6 \\ 

\\
Fe line energy (keV)             &6.39$\pm$0.09          &6.32$\pm$0.07   &6.37$\pm$0.07  &6.3$\pm$0.1    \\
Fe line eq. width (eV)           &71$\pm$10              &72$\pm$14       &88$\pm$18      &83$\pm$18   \\

\\
Cyclotron line energy (E$_{c}$) (keV)         &26.8$\pm$0.8    &28.1$^{+1.7}_{-1.4}$      &29.5$^{+2.7}_{-1.7}$ &31.5$^{+1.3}_{-1.1}$\\
Cyclotron line width ($\sigma_{c}$) (keV)     &5.5$\pm$0.8     &5.5$^{+1.5}_{-1.1}$       &6.6$^{+2.3}_{-1.4}$ &6.5$^{+0.9}_{-0.8}$\\
Cyclotron line strength ($\tau_{c}$)	      &3.9$^{+1.1}_{-0.8}$         &4.4$^{+2.9}_{-1.6}$  &6.2$^{+6.9}_{-2.5}$ &8.7$^{+3.1}_{-2.2}$\\
\\
Luminosity$^b$ (0.5-10 keV)   &2.75$\pm$0.02     &1.13$\pm$0.01    &0.7$\pm$0.01      &1.10$\pm$0.01 \\ 
Luminosity$^b$ (0.5-100 keV)   &6.08$\pm$0.02     &2.87$\pm$0.02    &1.97$\pm$0.01      &2.82$\pm$0.01 \\ 
$\chi^2_\nu$ ($\nu$)    &1.04 (1276)          &0.96 (1125)      &1.06 (602)        &1.09 (1549)  \\
\hline
\end{tabular}
\\
\flushleft
$^a$ : Equivalent hydrogen column density (in 10$^{22}$ atoms cm$^{-2}$); 
$^b$ : Luminosity in 10$^{38}$  ergs s$^{-1}$ units, assuming the source distance of 62~kpc \citep{Hilditch2005MNRAS.357..304H, Haschke2012AJ....144..107H, Graczyk2014ApJ...780...59G}. 
\\
\label{spec_par}
\end{table*}


\section{Discussion and Conclusions}

The broadband emission from the accreting X-ray pulsars is understood to be due to thermal and bulk Comptonizations of seed photons from the hot spots within the accretion column mounted on the magnetic poles of the neutron star \citep{Becker2007}. The underlying physical phenomena of these sources can be understood by studying the pulse emission geometry and the spectral evolution of the pulsar during X-ray outbursts. At lower luminosities such as 10$^{34-35}$ erg~s$^{-1}$, the accreted material falls freely until it gets halted and settled by Coulomb interactions in a hydrodynamical shock near the surface \citep{Basko1975}. This leads to a simplistic pulsed emission geometry in the form of a pencil beam where the X-ray photons propagate along the magnetic field lines. At this stage, the radiation is mostly dominated by bulk Comptonization \citep{Becker2005ApJ...630..465B,Becker2007}. When the accretion rate increases, a radiation-dominated shock is expected to be formed in the accretion column at a critical luminosity that leads to geometrical changes or spectral transition in the emission \citep{Basko1976, Becker2007}. The infalling material interacts closely with the shock and gets decelerated through it before settling down onto the neutron star. The photons below the shock region diffuse through the column side-wall in the form of a fan beam pattern. The anticipated emission geometry is a mixture of fan and pencil beam patterns at a luminosity of $\sim$10$^{36-37}$ erg~s$^{-1}$. The thermal as well as bulk Comptonization processes contribute significantly to the observed broadband emission at this stage. In the super-critical luminosity regime ($>$10$^{37}$ erg~s$^{-1}$), a pure fan beam pattern is expected because of a strong radiative shock that makes the accretion column optically thick for photons propagating along the magnetic field lines.

\source was observed with \nustar at luminosities above the Eddington limit of a classical neutron star during the 2015 and 2022 outbursts. The \nustar pulse profiles are found to be luminosity dependent. At a luminosity of 6.1$\times$10$^{38}$~ergs~s$^{-1}$ (Obs-I), a doubled peaked profile appears throughout soft to hard X-rays suggesting the emission from both the poles of the neutron star as the observed peaks are separated by 0.5 phase in the pulse profile. The emission geometry, however, evolves below this luminosity, where the soft X-ray profiles are mainly broad or contain a hump-like structure. The emission from both poles is clearly apparent in the hard X-rays in the case of Obs-II and IV.  The pulse fraction of the pulsar increases with energy clearly up to 40 keV, indicating an increase in the pulsating photons contributing to  emission at higher luminosities. This suggests the rising height of the accretion column or mound. The profile changes  with luminosity can be understood in terms of evolving emission geometry, where the significant contribution of the fan beam is anticipated  in the super-critical regime.  

We also detected a spectral transition in \source during the 2022 outburst with \nicer. The photon index shows a positive and negative dependency on luminosity below and above a range of value  in the range of (6--8)$\times$10$^{37}$~ergs~s$^{-1}$ in 0.5--10 keV range. Considering a bolometric correction factor of 2.5, we expect a critical luminosity of (1.5--2)$\times$10$^{38}$~ergs~s$^{-1}$ at which the emission pattern is expected to change. According to \citet{Becker2012A&A...544A.123B}, the critical luminosity depends on the magnetic field assuming disk accretion onto a classical neutron star with a 1.4\msol mass and a 10 km radius.

\begin{eqnarray}
	L_{\rm crit} &=& 1.49 \times 10^{37} \left( \frac{B}{10^{12}{\rm\,G}} \right)^{16/15} {\rm erg\,s}^{-1}
	\label{eqn:lcrit-simplify}
\end{eqnarray}

Using equation~\ref{eqn:lcrit-simplify}, the magnetic field $B$ can be estimated to be in the range of (8.7--11.4)$\times$ 10$^{12}$~G for the above range of luminosity values. The estimated magnetic field differs by a factor of 3 to 4 from the measurement based on the detection of a cyclotron line in the spectrum \citep{Jaisawal2016MNRAS.461L..97J}. 
 In case no bolometric correction is applied, the magnetic field from equation~\ref{eqn:lcrit-simplify} turns out to be  (3.7--4.8)$\times$ 10$^{12}$~G i.e., much closer to the estimates based on cyclotron line energy.

\subsection{The evolution of cyclotron absorption line and its implication on the magnetic field}

Both the 2015 and 2022 X-ray outbursts of \source lasted for more than 2 months with peak bolometric luminosities of 6$\times$10$^{38}$~ergs~s$^{-1}$ and 3$\times$10$^{38}$~ergs~s$^{-1}$, respectively (see Figure~\ref{outburst_lc}). We report broadband spectral properties of the pulsar during the latest giant outburst using data from \nicer and \nustar observations. The 2015 \nustar and {\it Swift}/XRT observations are also used to study the cyclotron absorption line and its long-term evolution. We found relatively higher cyclotron line energy (with an increase of 3.4 keV) during the current outburst in comparison to the 2015 measurement made around a similar luminosity. In addition, the line energy is anti-correlated with luminosity in the 2015 outburst. The cyclotron line is a unique absorption-like feature that is observed in the 10-100 keV spectrum of X-ray pulsars \citep{Meszaros1992herm.book.....M}. They appear due to a resonance scattering of photons with electrons in quantized Landau energy states in the presence of a strong magnetic field. The field strength can be estimated directly using a 12-B-12 rule depending on the line energy since it is expressed as {\it E$_{cyc}$=11.6B$_{12}\times(1+z){^{-1}}$} keV, where  B$_{12}$ is the magnetic field in 10$^{12}$~G unit, and $z$ is the gravitational red-shift. 

The centroid energy of the cyclotron line in principle measures the field strength at the line-forming region. A change in line energy with luminosity is observed in several accretion-powered X-ray pulsars during outbursts \citep{Nakajima2006ApJ...646.1125N, Tsygankov2010MNRAS.401.1628T, Jaisawal2016MNRAS.457.2749J, Staubert2019}. In some cases like Her~X-1 \citep{Staubert2007A&A...465L..25S}, the cyclotron energy and luminosity are positively correlated, whereas there are pulsars such as 4U~0115+63 \citep{Nakajima2006ApJ...646.1125N, Tsygankov2010MNRAS.401.1628T} and V~0332+53 \citep{Tsygankov2010MNRAS.401.1628T} that show a negative correlation between these parameters \citep{Becker2012A&A...544A.123B}. The positive correlation is expected in the sub-critical region (below the critical luminosity) when the pressure of the accreting plasma pushes the hydrodynamical shock or line-forming region closer to the neutron star surface with an increasing accretion rate (luminosity). On the other hand, a formation of radiation-dominated shock is expected in the accretion column at higher accretion rates. The shock drifts upwards in the column due to strong radiation pressure with increasing mass accretion rate in the super-critical regime \citep{Becker2012A&A...544A.123B}. Therefore, a negative correlation between the cyclotron line energy and luminosity is expected at this phase. This is observed in SMC~X-2 when the pulsar was accreting in the super-critical regime during the 2015 outburst. Alternatively, the formation of a cyclotron line and its observed negative correlation can be understood by the reflection of accretion column photons onto the neutron star surface \citep{Poutanen2013ApJ...777..115P}. The hypothesis suggests that the cyclotron line gets formed on the surface rather than in the accretion column where the gradient is high to explain the observed changes in the line energy. As the accretion rate increases, a larger accretion column covers a substantial portion of the atmosphere from the pole to the equator (high to low magnetic field regions). This can also explain the origin of the observed negative correlation between the cyclotron energy and luminosity in SMC~X-2 resulted due to a change in illumination pattern.

It is important to note that the  cyclotron line energy of \source is different at a given luminosity observed at Obs-II and Obs-IV during different outbursts. The \maxi/GSC light curves in Figure~\ref{outburst_lc} show a clear rise in the source intensity after MJD 57259 and 59738 during the 2015 and 2022 outbursts, respectively. The source reached a luminosity of (2--3)$\times$10$^{38}$~ergs~s$^{-1}$ (Obs-II and IV) after 49 and 36 days since the onset of respective outbursts. Our measurements suggest an increase of 3.4 keV in the cyclotron line energy between these data sets at comparable luminosities. This difference in the cyclotron line energy corresponds to a change in the magnetic field of 3.8$\times$10$^{11}$~G, assuming a classical neutron star and gravitational redshift of 0.3. Such an enhancement in the cyclotron energy can be understood by an accretion-induced screening of the magnetic field or geometrical changes in the line-forming region. Theoretical studies suggest that the accretion flow distribution from poles to lower latitudes may drag the field lines (\citealt{Choudhuri2002MNRAS.332..933C} and references therein). This mechanism could effectively lower down the original field strength even on a shorter time scale in the top layer of the distributed material, considering no effect of magnetic buoyancy. However, in reality, the screening mechanism depends on the ratio of timescales for field burial by the advection mechanism and the re-emerging of the field through magnetic buoyancy and Ohmic rediffusion \citep{Choudhuri2002MNRAS.332..933C}. An increase of 3.4 keV in cyclotron line energy in the latest modest outburst suggests that the screening mechanism is not effective as it is in the case of the 2015 giant outburst, where a similar luminosity is achieved in the decay phase almost 49 days after the onset. The neutron star by this point would accumulate a factor of two or more mass with an integrated value of 3.5$\times$10$^{24}$~g in 2015. 

A similar possibility is discussed for V~0332+53 during its 2004/05 outburst where the cyclotron line energy at a given luminosity in the decay  got reduced by $\approx$1.5 keV with respect to the rising phase of the outburst \citep{Cusumano2016MNRAS.460L..99C}. However,  \citet{Doroshenko2017MNRAS.466.2143D} argued that the expected time scales for field burring/reemerging are significantly shorter than those observed in the case of Vela~X-1 and Her~X-1. They also noticed discrepancies between the spin rates in rising and declining parts of the outburst that could explain the difference in the magnetic field based on the accretion-torque theory. In addition, they suggested that the change in the accretion disk and effective magnetospheric radius in different phases of the outburst lead to a variable column height, resulting in an evolving cyclotron line through X-ray reflection on the surface \citep{Doroshenko2017MNRAS.466.2143D}. In the case of \source, however, we observe the increase in cyclotron line energy in the decay phase of two outbursts at a comparable luminosity. This is important to recall that the pulse profile observed during Obs-II and IV appears identical, suggesting similar accretion geometry of the pulsar. The accretion-induced screening mechanism, thus, seems a possible explanation for the observed behavior over the geometrical change in the line-forming region.

The long term evolution of the cyclotron line is well studied for two sources, Her~X-1 and Vela X-1 \citep{LaParola2016MNRAS.463..185L, Staubert2019}. The cyclotron centroid energy of Her~X-1 has shown a secular decay of $\approx$5~keV in 20 years before stabilizing between 2012 and 2019. There was also a period in 1993-1996 when the line energy was increased by a few keV. The time evolution of the cyclotron line energy is still a matter of study. It is believed that the observed decay/increase in line energy represents a geometrical change within the emission region or in the configuration of the local magnetic field rather than a change in the global dipolar magnetic field of the neutron star \citep{Mukherjee2013MNRAS.430.1976M,Staubert2019}. This could be further understood in terms of a marginal imbalance between the accretion rate and the losing or settling rate of the accreted material on the accretion mound. If the settling rate is slightly higher than the leakage of the material from the bottom of the mound, this may lead to an increase in the column height as well as a change in the local magnetic field configuration of the line forming region. A line decay is expected naturally in this case. Over time, the excessive accumulation of the material and its pressure on the mound may result in a forced redistribution of the material across the surface. This would adjust the accretion mount to its unperturbed configuration where higher line energy would be expected. The time scale for such redistribution is unclear. But, it may happen after continuous accumulation of matter over a long period or (multiple) strong outbursts like those observed in 2002 and 2015 of \source. Future observations of the source during outbursts will allow understanding the behavior of the cyclotron line better.

\

\subsection*{ACKNOWLEDGMENTS}
We thank the referee for constructive suggestions on the paper.
This research has made use of data obtained through 
HEASARC Online Service, provided by the NASA/GSFC, in support of NASA 
High Energy Astrophysics Programs. This work used the NuSTAR Data Analysis 
Software ({\tt NuSTARDAS}) jointly developed by the ASI
Science Data Center (ASDC, Italy) and the California Institute of Technology (USA).
SG acknowledges the support of the CNES. GV acknowledges support by 
H.F.R.I. through the project ASTRAPE (Project ID 7802).

\section*{DATA AVAILABILITY}
We used archival data of {\it NICER}, {\it NuSTAR}, {\it Swift}, and {\it MAXI} observatories for this work. 

\bibliographystyle{mnras}
\bibliography{references.bib}

\begin{thebibliography}{}
\makeatletter
\relax
\def\mn@urlcharsother{\let\do\@makeother \do\$\do\&\do\#\do\^\do\_\do\%\do\~}
\def\mn@doi{\begingroup\mn@urlcharsother \@ifnextchar [ {\mn@doi@}
  {\mn@doi@[]}}
\def\mn@doi@[#1]#2{\def\@tempa{#1}\ifx\@tempa\@empty \href
  {http://dx.doi.org/#2} {doi:#2}\else \href {http://dx.doi.org/#2} {#1}\fi
  \endgroup}
\def\mn@eprint#1#2{\mn@eprint@#1:#2::\@nil}
\def\mn@eprint@arXiv#1{\href {http://arxiv.org/abs/#1} {{\tt arXiv:#1}}}
\def\mn@eprint@dblp#1{\href {http://dblp.uni-trier.de/rec/bibtex/#1.xml}
  {dblp:#1}}
\def\mn@eprint@#1:#2:#3:#4\@nil{\def\@tempa {#1}\def\@tempb {#2}\def\@tempc
  {#3}\ifx \@tempc \@empty \let \@tempc \@tempb \let \@tempb \@tempa \fi \ifx
  \@tempb \@empty \def\@tempb {arXiv}\fi \@ifundefined
  {mn@eprint@\@tempb}{\@tempb:\@tempc}{\expandafter \expandafter \csname
  mn@eprint@\@tempb\endcsname \expandafter{\@tempc}}}

\bibitem[\protect\citeauthoryear{{Anastasopoulou}, {Zezas}, {Steiner}  \&
  {Reig}}{{Anastasopoulou} et~al.}{2022}]{2022MNRAS.513.1400A}
{Anastasopoulou} K.,  {Zezas} A.,  {Steiner} J.~F.,   {Reig} P.,  2022, \mn@doi
  [\mnras] {10.1093/mnras/stac940}, \href
  {https://ui.adsabs.harvard.edu/abs/2022MNRAS.513.1400A} {513, 1400}

\bibitem[\protect\citeauthoryear{{Arnaud}}{{Arnaud}}{1996}]{Arnaud1996}
{Arnaud} K.~A.,  1996, in {Jacoby} G.~H.,  {Barnes} J.,  eds,  Astronomical
  Society of the Pacific Conference Series Vol. 101, Astronomical Data Analysis
  Software and Systems V. p.~17

\bibitem[\protect\citeauthoryear{{Basko} \& {Sunyaev}}{{Basko} \&
  {Sunyaev}}{1975}]{Basko1975}
{Basko} M.~M.,  {Sunyaev} R.~A.,  1975, Astronomy and Astrophysics, \href
  {https://ui.adsabs.harvard.edu/abs/1975A&A....42..311B} {42, 311}

\bibitem[\protect\citeauthoryear{{Basko} \& {Sunyaev}}{{Basko} \&
  {Sunyaev}}{1976}]{Basko1976}
{Basko} M.~M.,  {Sunyaev} R.~A.,  1976, \mn@doi [\mnras]
  {10.1093/mnras/175.2.395}, \href
  {http://adsabs.harvard.edu/abs/1976MNRAS.175..395B} {175, 395}

\bibitem[\protect\citeauthoryear{{Becker} \& {Wolff}}{{Becker} \&
  {Wolff}}{2005}]{Becker2005ApJ...630..465B}
{Becker} P.~A.,  {Wolff} M.~T.,  2005, \mn@doi [\apj] {10.1086/431720}, \href
  {https://ui.adsabs.harvard.edu/abs/2005ApJ...630..465B} {630, 465}

\bibitem[\protect\citeauthoryear{{Becker} \& {Wolff}}{{Becker} \&
  {Wolff}}{2007}]{Becker2007}
{Becker} P.~A.,  {Wolff} M.~T.,  2007, \mn@doi [\apj] {10.1086/509108}, \href
  {https://ui.adsabs.harvard.edu/abs/2007ApJ...654..435B} {654, 435}

\bibitem[\protect\citeauthoryear{{Becker} et~al.,}{{Becker}
  et~al.}{2012}]{Becker2012A&A...544A.123B}
{Becker} P.~A.,  et~al., 2012, \mn@doi [\aap] {10.1051/0004-6361/201219065},
  \href {https://ui.adsabs.harvard.edu/abs/2012A&A...544A.123B} {544, A123}

\bibitem[\protect\citeauthoryear{{Choudhuri} \& {Konar}}{{Choudhuri} \&
  {Konar}}{2002}]{Choudhuri2002MNRAS.332..933C}
{Choudhuri} A.~R.,  {Konar} S.,  2002, \mn@doi [\mnras]
  {10.1046/j.1365-8711.2002.05362.x}, \href
  {https://ui.adsabs.harvard.edu/abs/2002MNRAS.332..933C} {332, 933}

\bibitem[\protect\citeauthoryear{{Clark}, {Doxsey}, {Li}, {Jernigan}  \& {van
  Paradijs}}{{Clark} et~al.}{1978}]{Clark1978ApJ...221L..37C}
{Clark} G.,  {Doxsey} R.,  {Li} F.,  {Jernigan} J.~G.,   {van Paradijs} J.,
  1978, \mn@doi [\apjl] {10.1086/182660}, \href
  {https://ui.adsabs.harvard.edu/abs/1978ApJ...221L..37C} {221, L37}

\bibitem[\protect\citeauthoryear{{Clark}, {Li}  \& {van Paradijs}}{{Clark}
  et~al.}{1979}]{Clark1979ApJ...227...54C}
{Clark} G.,  {Li} F.,   {van Paradijs} J.,  1979, \mn@doi [\apj]
  {10.1086/156702}, \href
  {https://ui.adsabs.harvard.edu/abs/1979ApJ...227...54C} {227, 54}

\bibitem[\protect\citeauthoryear{{Coe} \& {Kirk}}{{Coe} \&
  {Kirk}}{2015}]{Coe2015MNRAS.452..969C}
{Coe} M.~J.,  {Kirk} J.,  2015, \mn@doi [\mnras] {10.1093/mnras/stv1283}, \href
  {https://ui.adsabs.harvard.edu/abs/2015MNRAS.452..969C} {452, 969}

\bibitem[\protect\citeauthoryear{{Coe}, {Kennea}, {Evans}, {Buckley},
  {Townsend}  \& {Monageng}}{{Coe} et~al.}{2022}]{Coe2022ATel15500....1C}
{Coe} M.~J.,  {Kennea} J.~A.,  {Evans} P.~A.,  {Buckley} D.~A.~H.,  {Townsend}
  L.~J.,   {Monageng} I.,  2022, The Astronomer's Telegram, \href
  {https://ui.adsabs.harvard.edu/abs/2022ATel15500....1C} {15500, 1}

\bibitem[\protect\citeauthoryear{{Corbet}, {Marshall}, {Coe}, {Laycock}  \&
  {Handler}}{{Corbet} et~al.}{2001}]{Corbet2001ApJ...548L..41C}
{Corbet} R.~H.~D.,  {Marshall} F.~E.,  {Coe} M.~J.,  {Laycock} S.,   {Handler}
  G.,  2001, \mn@doi [\apjl] {10.1086/318929}, \href
  {https://ui.adsabs.harvard.edu/abs/2001ApJ...548L..41C} {548, L41}

\bibitem[\protect\citeauthoryear{{Cusumano}, {La Parola}, {D'A{\`\i}},
  {Segreto}, {Tagliaferri}, {Barthelmy}  \& {Gehrels}}{{Cusumano}
  et~al.}{2016}]{Cusumano2016MNRAS.460L..99C}
{Cusumano} G.,  {La Parola} V.,  {D'A{\`\i}} A.,  {Segreto} A.,  {Tagliaferri}
  G.,  {Barthelmy} S.~D.,   {Gehrels} N.,  2016, \mn@doi [\mnras]
  {10.1093/mnrasl/slw084}, \href
  {https://ui.adsabs.harvard.edu/abs/2016MNRAS.460L..99C} {460, L99}

\bibitem[\protect\citeauthoryear{{Dickey} \& {Lockman}}{{Dickey} \&
  {Lockman}}{1990}]{Dickey1990ARA&A..28..215D}
{Dickey} J.~M.,  {Lockman} F.~J.,  1990, \mn@doi [\araa]
  {10.1146/annurev.aa.28.090190.001243}, \href
  {https://ui.adsabs.harvard.edu/abs/1990ARA&A..28..215D} {28, 215}

\bibitem[\protect\citeauthoryear{{Doroshenko}, {Tsygankov}, {Mushtukov},
  {Lutovinov}, {Santangelo}, {Suleimanov}  \& {Poutanen}}{{Doroshenko}
  et~al.}{2017}]{Doroshenko2017MNRAS.466.2143D}
{Doroshenko} V.,  {Tsygankov} S.~S.,  {Mushtukov} A.~A.,  {Lutovinov} A.~A.,
  {Santangelo} A.,  {Suleimanov} V.~F.,   {Poutanen} J.,  2017, \mn@doi
  [\mnras] {10.1093/mnras/stw3236}, \href
  {https://ui.adsabs.harvard.edu/abs/2017MNRAS.466.2143D} {466, 2143}

\bibitem[\protect\citeauthoryear{{Evans} et~al.,}{{Evans}
  et~al.}{2009}]{Evans2009}
{Evans} P.~A.,  et~al., 2009, \mn@doi [\mnras]
  {10.1111/j.1365-2966.2009.14913.x}, \href
  {http://adsabs.harvard.edu/abs/2009MNRAS.397.1177E} {397, 1177}

\bibitem[\protect\citeauthoryear{{Gendreau}, {Arzoumanian}  \&
  {Okajima}}{{Gendreau} et~al.}{2012}]{Gendreau2012}
{Gendreau} K.~C.,  {Arzoumanian} Z.,   {Okajima} T.,  2012, in Society of
  Photo-Optical Instrumentation Engineers (SPIE) Conference Series. p.~13,
  \mn@doi{10.1117/12.926396}

\bibitem[\protect\citeauthoryear{{Gendreau} et~al.}{{Gendreau}
  et~al.}{2016}]{Gendreau2016}
{Gendreau} K.~C.,  et~al., 2016, in Space Telescopes and Instrumentation 2016:
  Ultraviolet to Gamma Ray. p. 99051H, \mn@doi{10.1117/12.2231304}

\bibitem[\protect\citeauthoryear{{Graczyk} et~al.,}{{Graczyk}
  et~al.}{2014}]{Graczyk2014ApJ...780...59G}
{Graczyk} D.,  et~al., 2014, \mn@doi [\apj] {10.1088/0004-637X/780/1/59}, \href
  {https://ui.adsabs.harvard.edu/abs/2014ApJ...780...59G} {780, 59}

\bibitem[\protect\citeauthoryear{{Haberl} \& {Sturm}}{{Haberl} \&
  {Sturm}}{2016}]{Haberl2016}
{Haberl} F.,  {Sturm} R.,  2016, \mn@doi [\aap] {10.1051/0004-6361/201527326},
  \href {https://ui.adsabs.harvard.edu/abs/2016A&A...586A..81H} {586, A81}

\bibitem[\protect\citeauthoryear{{Harrison} et~al.,}{{Harrison}
  et~al.}{2013}]{Harrison2013}
{Harrison} F.~A.,  et~al., 2013, \mn@doi [\apj] {10.1088/0004-637X/770/2/103},
  \href {http://adsabs.harvard.edu/abs/2013ApJ...770..103H} {770, 103}

\bibitem[\protect\citeauthoryear{{Haschke}, {Grebel}  \& {Duffau}}{{Haschke}
  et~al.}{2012}]{Haschke2012AJ....144..107H}
{Haschke} R.,  {Grebel} E.~K.,   {Duffau} S.,  2012, \mn@doi [\aj]
  {10.1088/0004-6256/144/4/107}, \href
  {https://ui.adsabs.harvard.edu/abs/2012AJ....144..107H} {144, 107}

\bibitem[\protect\citeauthoryear{{Hilditch}, {Howarth}  \&
  {Harries}}{{Hilditch} et~al.}{2005}]{Hilditch2005MNRAS.357..304H}
{Hilditch} R.~W.,  {Howarth} I.~D.,   {Harries} T.~J.,  2005, \mn@doi [\mnras]
  {10.1111/j.1365-2966.2005.08653.x}, \href
  {https://ui.adsabs.harvard.edu/abs/2005MNRAS.357..304H} {357, 304}

\bibitem[\protect\citeauthoryear{{Huppenkothen} et~al.,}{{Huppenkothen}
  et~al.}{2019}]{2019ApJ...881...39H}
{Huppenkothen} D.,  et~al., 2019, \mn@doi [\apj] {10.3847/1538-4357/ab258d},
  \href {https://ui.adsabs.harvard.edu/abs/2019ApJ...881...39H} {881, 39}

\bibitem[\protect\citeauthoryear{{Jaisawal} \& {Naik}}{{Jaisawal} \&
  {Naik}}{2016}]{Jaisawal2016MNRAS.461L..97J}
{Jaisawal} G.~K.,  {Naik} S.,  2016, \mn@doi [\mnras] {10.1093/mnrasl/slw108},
  \href {https://ui.adsabs.harvard.edu/abs/2016MNRAS.461L..97J} {461, L97}

\bibitem[\protect\citeauthoryear{{Jaisawal}, {Naik}  \& {Epili}}{{Jaisawal}
  et~al.}{2016}]{Jaisawal2016MNRAS.457.2749J}
{Jaisawal} G.~K.,  {Naik} S.,   {Epili} P.,  2016, \mn@doi [\mnras]
  {10.1093/mnras/stw085}, \href
  {https://ui.adsabs.harvard.edu/abs/2016MNRAS.457.2749J} {457, 2749}

\bibitem[\protect\citeauthoryear{{Kahabka} \& {Pietsch}}{{Kahabka} \&
  {Pietsch}}{1996}]{Kahabka1996A&A...312..919K}
{Kahabka} P.,  {Pietsch} W.,  1996, \aap, \href
  {https://ui.adsabs.harvard.edu/abs/1996A&A...312..919K} {312, 919}

\bibitem[\protect\citeauthoryear{{Karaferias}, {Vasilopoulos}, {Petropoulou},
  {Jenke}, {Wilson-Hodge}  \& {Malacaria}}{{Karaferias}
  et~al.}{2023}]{Karaferias2023MNRAS.520..281K}
{Karaferias} A.~S.,  {Vasilopoulos} G.,  {Petropoulou} M.,  {Jenke} P.~A.,
  {Wilson-Hodge} C.~A.,   {Malacaria} C.,  2023, \mn@doi [\mnras]
  {10.1093/mnras/stac3208}, \href
  {https://ui.adsabs.harvard.edu/abs/2023MNRAS.520..281K} {520, 281}

\bibitem[\protect\citeauthoryear{{Kennea} et~al.,}{{Kennea}
  et~al.}{2015}]{Kennea2015ATel.8091....1K}
{Kennea} J.~A.,  et~al., 2015, The Astronomer's Telegram, \href
  {https://ui.adsabs.harvard.edu/abs/2015ATel.8091....1K} {8091, 1}

\bibitem[\protect\citeauthoryear{{Kennea}, {Coe}  \& {Evans}}{{Kennea}
  et~al.}{2022}]{Kennea2022ATel15434....1K}
{Kennea} J.~A.,  {Coe} M.~J.,   {Evans} P.~A.,  2022, The Astronomer's
  Telegram, \href {https://ui.adsabs.harvard.edu/abs/2022ATel15434....1K}
  {15434, 1}

\bibitem[\protect\citeauthoryear{{La Palombara}, {Sidoli}, {Pintore},
  {Esposito}, {Mereghetti}  \& {Tiengo}}{{La Palombara}
  et~al.}{2016}]{Palombara2016MNRAS.458L..74L}
{La Palombara} N.,  {Sidoli} L.,  {Pintore} F.,  {Esposito} P.,  {Mereghetti}
  S.,   {Tiengo} A.,  2016, \mn@doi [\mnras] {10.1093/mnrasl/slw020}, \href
  {https://ui.adsabs.harvard.edu/abs/2016MNRAS.458L..74L} {458, L74}

\bibitem[\protect\citeauthoryear{{La Parola}, {Cusumano}, {Segreto}  \&
  {D'A{\`\i}}}{{La Parola} et~al.}{2016}]{LaParola2016MNRAS.463..185L}
{La Parola} V.,  {Cusumano} G.,  {Segreto} A.,   {D'A{\`\i}} A.,  2016, \mn@doi
  [\mnras] {10.1093/mnras/stw1915}, \href
  {https://ui.adsabs.harvard.edu/abs/2016MNRAS.463..185L} {463, 185}

\bibitem[\protect\citeauthoryear{{Leahy}}{{Leahy}}{1987}]{Leahy1987}
{Leahy} D.~A.,  1987, \aap, \href
  {https://ui.adsabs.harvard.edu/abs/1987A&A...180..275L} {180, 275}

\bibitem[\protect\citeauthoryear{{Li}, {Hu}, {Lin}  \& {Kong}}{{Li}
  et~al.}{2016}]{Li2016ApJ...828...74L}
{Li} K.~L.,  {Hu} C.~P.,  {Lin} L.~C.~C.,   {Kong} A. K.~H.,  2016, \mn@doi
  [\apj] {10.3847/0004-637X/828/2/74}, \href
  {https://ui.adsabs.harvard.edu/abs/2016ApJ...828...74L} {828, 74}

\bibitem[\protect\citeauthoryear{{Liu}, {van Paradijs}  \& {van den
  Heuvel}}{{Liu} et~al.}{2006}]{Liu2006}
{Liu} Q.~Z.,  {van Paradijs} J.,   {van den Heuvel} E.~P.~J.,  2006, \mn@doi
  [\aap] {10.1051/0004-6361:20064987}, \href
  {https://ui.adsabs.harvard.edu/abs/2006A&A...455.1165L} {455, 1165}

\bibitem[\protect\citeauthoryear{{Luo} et~al.,}{{Luo}
  et~al.}{2021}]{2021ApJ...911...45L}
{Luo} J.,  et~al., 2021, \mn@doi [\apj] {10.3847/1538-4357/abe62f}, \href
  {https://ui.adsabs.harvard.edu/abs/2021ApJ...911...45L} {911, 45}

\bibitem[\protect\citeauthoryear{{Lutovinov}, {Tsygankov}, {Krivonos}, {Molkov}
   \& {Poutanen}}{{Lutovinov} et~al.}{2017}]{Lutovinov2017ApJ...834..209L}
{Lutovinov} A.~A.,  {Tsygankov} S.~S.,  {Krivonos} R.~A.,  {Molkov} S.~V.,
  {Poutanen} J.,  2017, \mn@doi [\apj] {10.3847/1538-4357/834/2/209}, \href
  {https://ui.adsabs.harvard.edu/abs/2017ApJ...834..209L} {834, 209}

\bibitem[\protect\citeauthoryear{{Makishima}, {Mihara}, {Nagase}  \&
  {Tanaka}}{{Makishima} et~al.}{1999}]{Makishima1999ApJ...525..978M}
{Makishima} K.,  {Mihara} T.,  {Nagase} F.,   {Tanaka} Y.,  1999, \mn@doi
  [\apj] {10.1086/307912}, \href
  {https://ui.adsabs.harvard.edu/abs/1999ApJ...525..978M} {525, 978}

\bibitem[\protect\citeauthoryear{{Marshall}, {Boldt}, {Holt}, {Mushotzky},
  {Pravdo}, {Rothschild}  \& {Serlemitsos}}{{Marshall}
  et~al.}{1979}]{Marshall1979ApJS...40..657M}
{Marshall} F.~E.,  {Boldt} E.~A.,  {Holt} S.~S.,  {Mushotzky} R.~F.,  {Pravdo}
  S.~H.,  {Rothschild} R.~E.,   {Serlemitsos} P.~J.,  1979, \mn@doi [\apjs]
  {10.1086/190600}, \href
  {https://ui.adsabs.harvard.edu/abs/1979ApJS...40..657M} {40, 657}

\bibitem[\protect\citeauthoryear{{Matsuoka} et~al.,}{{Matsuoka}
  et~al.}{2009}]{Matsuoka2009PASJ...61..999M}
{Matsuoka} M.,  et~al., 2009, \mn@doi [\pasj] {10.1093/pasj/61.5.999}, \href
  {https://ui.adsabs.harvard.edu/abs/2009PASJ...61..999M} {61, 999}

\bibitem[\protect\citeauthoryear{{McBride}, {Coe}, {Negueruela}, {Schurch}  \&
  {McGowan}}{{McBride} et~al.}{2008}]{McBride2008MNRAS.388.1198M}
{McBride} V.~A.,  {Coe} M.~J.,  {Negueruela} I.,  {Schurch} M.~P.~E.,
  {McGowan} K.~E.,  2008, \mn@doi [\mnras] {10.1111/j.1365-2966.2008.13410.x},
  \href {https://ui.adsabs.harvard.edu/abs/2008MNRAS.388.1198M} {388, 1198}

\bibitem[\protect\citeauthoryear{{Meszaros}}{{Meszaros}}{1992}]{Meszaros1992herm.book.....M}
{Meszaros} P.,  1992, {High-energy radiation from magnetized neutron stars}

\bibitem[\protect\citeauthoryear{{Mukherjee}, {Bhattacharya}  \&
  {Mignone}}{{Mukherjee} et~al.}{2013}]{Mukherjee2013MNRAS.430.1976M}
{Mukherjee} D.,  {Bhattacharya} D.,   {Mignone} A.,  2013, \mn@doi [\mnras]
  {10.1093/mnras/stt020}, \href
  {https://ui.adsabs.harvard.edu/abs/2013MNRAS.430.1976M} {430, 1976}

\bibitem[\protect\citeauthoryear{{Nakajima}, {Mihara}, {Makishima}  \&
  {Niko}}{{Nakajima} et~al.}{2006}]{Nakajima2006ApJ...646.1125N}
{Nakajima} M.,  {Mihara} T.,  {Makishima} K.,   {Niko} H.,  2006, \mn@doi
  [\apj] {10.1086/502638}, \href
  {https://ui.adsabs.harvard.edu/abs/2006ApJ...646.1125N} {646, 1125}

\bibitem[\protect\citeauthoryear{{Negoro} et~al.,}{{Negoro}
  et~al.}{2015}]{Negoro2015ATel.8088....1N}
{Negoro} H.,  et~al., 2015, The Astronomer's Telegram, \href
  {https://ui.adsabs.harvard.edu/abs/2015ATel.8088....1N} {8088, 1}

\bibitem[\protect\citeauthoryear{{Porter} \& {Rivinius}}{{Porter} \&
  {Rivinius}}{2003}]{Porter2003}
{Porter} J.~M.,  {Rivinius} T.,  2003, \mn@doi [\pasp] {10.1086/378307}, \href
  {https://ui.adsabs.harvard.edu/abs/2003PASP..115.1153P} {115, 1153}

\bibitem[\protect\citeauthoryear{{Poutanen}, {Mushtukov}, {Suleimanov},
  {Tsygankov}, {Nagirner}, {Doroshenko}  \& {Lutovinov}}{{Poutanen}
  et~al.}{2013}]{Poutanen2013ApJ...777..115P}
{Poutanen} J.,  {Mushtukov} A.~A.,  {Suleimanov} V.~F.,  {Tsygankov} S.~S.,
  {Nagirner} D.~I.,  {Doroshenko} V.,   {Lutovinov} A.~A.,  2013, \mn@doi
  [\apj] {10.1088/0004-637X/777/2/115}, \href
  {https://ui.adsabs.harvard.edu/abs/2013ApJ...777..115P} {777, 115}

\bibitem[\protect\citeauthoryear{{Prigozhin} et~al.,}{{Prigozhin}
  et~al.}{2012}]{Prigozhin2012}
{Prigozhin} G.,  et~al., 2012, in High Energy, Optical, and Infrared Detectors
  for Astronomy V. p. 845318, \mn@doi{10.1117/12.926667}

\bibitem[\protect\citeauthoryear{{Reig}}{{Reig}}{2011}]{Reig2011}
{Reig} P.,  2011, \mn@doi [\apss] {10.1007/s10509-010-0575-8}, \href
  {https://ui.adsabs.harvard.edu/abs/2011Ap&SS.332....1R} {332, 1}

\bibitem[\protect\citeauthoryear{{Remillard} et~al.,}{{Remillard}
  et~al.}{2022}]{Remillard2022AJ....163..130R}
{Remillard} R.~A.,  et~al., 2022, \mn@doi [\aj] {10.3847/1538-3881/ac4ae6},
  \href {https://ui.adsabs.harvard.edu/abs/2022AJ....163..130R} {163, 130}

\bibitem[\protect\citeauthoryear{{Roy}, {Cappallo}, {Laycock}, {Christodoulou},
  {Vasilopoulos}  \& {Bhattacharya}}{{Roy}
  et~al.}{2022}]{Roy2022ApJ...936...90R}
{Roy} A.,  {Cappallo} R.,  {Laycock} S. G.~T.,  {Christodoulou} D.~M.,
  {Vasilopoulos} G.,   {Bhattacharya} S.,  2022, \mn@doi [\apj]
  {10.3847/1538-4357/ac82b6}, \href
  {https://ui.adsabs.harvard.edu/abs/2022ApJ...936...90R} {936, 90}

\bibitem[\protect\citeauthoryear{{Russell} \& {Dopita}}{{Russell} \&
  {Dopita}}{1992}]{Russell1992ApJ...384..508R}
{Russell} S.~C.,  {Dopita} M.~A.,  1992, \mn@doi [\apj] {10.1086/170893}, \href
  {https://ui.adsabs.harvard.edu/abs/1992ApJ...384..508R} {384, 508}

\bibitem[\protect\citeauthoryear{{Schurch}, {Coe}, {McBride}, {Townsend},
  {Udalski}, {Haberl}  \& {Corbet}}{{Schurch}
  et~al.}{2011}]{Schurch2011MNRAS.412..391S}
{Schurch} M.~P.~E.,  {Coe} M.~J.,  {McBride} V.~A.,  {Townsend} L.~J.,
  {Udalski} A.,  {Haberl} F.,   {Corbet} R.~H.~D.,  2011, \mn@doi [\mnras]
  {10.1111/j.1365-2966.2010.17914.x}, \href
  {https://ui.adsabs.harvard.edu/abs/2011MNRAS.412..391S} {412, 391}

\bibitem[\protect\citeauthoryear{{Seward} \& {Mitchell}}{{Seward} \&
  {Mitchell}}{1981}]{Seward1981ApJ...243..736S}
{Seward} F.~D.,  {Mitchell} M.,  1981, \mn@doi [\apj] {10.1086/158641}, \href
  {https://ui.adsabs.harvard.edu/abs/1981ApJ...243..736S} {243, 736}

\bibitem[\protect\citeauthoryear{{Staubert}, {Shakura}, {Postnov}, {Wilms},
  {Rothschild}, {Coburn}, {Rodina}  \& {Klochkov}}{{Staubert}
  et~al.}{2007}]{Staubert2007A&A...465L..25S}
{Staubert} R.,  {Shakura} N.~I.,  {Postnov} K.,  {Wilms} J.,  {Rothschild}
  R.~E.,  {Coburn} W.,  {Rodina} L.,   {Klochkov} D.,  2007, \mn@doi [\aap]
  {10.1051/0004-6361:20077098}, \href
  {https://ui.adsabs.harvard.edu/abs/2007A&A...465L..25S} {465, L25}

\bibitem[\protect\citeauthoryear{{Staubert} et~al.,}{{Staubert}
  et~al.}{2019}]{Staubert2019}
{Staubert} R.,  et~al., 2019, \mn@doi [\aap] {10.1051/0004-6361/201834479},
  \href {https://ui.adsabs.harvard.edu/abs/2019A&A...622A..61S} {622, A61}

\bibitem[\protect\citeauthoryear{{Townsend}, {Coe}, {Corbet}  \&
  {Hill}}{{Townsend} et~al.}{2011}]{Townsend2011MNRAS.416.1556T}
{Townsend} L.~J.,  {Coe} M.~J.,  {Corbet} R.~H.~D.,   {Hill} A.~B.,  2011,
  \mn@doi [\mnras] {10.1111/j.1365-2966.2011.19153.x}, \href
  {https://ui.adsabs.harvard.edu/abs/2011MNRAS.416.1556T} {416, 1556}

\bibitem[\protect\citeauthoryear{{Tsygankov}, {Lutovinov}  \&
  {Serber}}{{Tsygankov} et~al.}{2010}]{Tsygankov2010MNRAS.401.1628T}
{Tsygankov} S.~S.,  {Lutovinov} A.~A.,   {Serber} A.~V.,  2010, \mn@doi
  [\mnras] {10.1111/j.1365-2966.2009.15791.x}, \href
  {https://ui.adsabs.harvard.edu/abs/2010MNRAS.401.1628T} {401, 1628}

\bibitem[\protect\citeauthoryear{{Vasilopoulos}, {Haberl}, {Carpano}  \&
  {Maitra}}{{Vasilopoulos} et~al.}{2018}]{2018A&A...620L..12V}
{Vasilopoulos} G.,  {Haberl} F.,  {Carpano} S.,   {Maitra} C.,  2018, \mn@doi
  [\aap] {10.1051/0004-6361/201833442}, \href
  {https://ui.adsabs.harvard.edu/abs/2018A&A...620L..12V} {620, L12}

\bibitem[\protect\citeauthoryear{{Vasilopoulos}, {Jaisawal}, {Maitra},
  {Haberl}, {Maggi}  \& {Karaferias}}{{Vasilopoulos}
  et~al.}{2022}]{2022A&A...664A.194V}
{Vasilopoulos} G.,  {Jaisawal} G.~K.,  {Maitra} C.,  {Haberl} F.,  {Maggi} P.,
   {Karaferias} A.~S.,  2022, \mn@doi [\aap] {10.1051/0004-6361/202243909},
  \href {https://ui.adsabs.harvard.edu/abs/2022A&A...664A.194V} {664, A194}

\bibitem[\protect\citeauthoryear{{Verner}, {Ferland}, {Korista}  \&
  {Yakovlev}}{{Verner} et~al.}{1996}]{Verner1996ApJ...465..487V}
{Verner} D.~A.,  {Ferland} G.~J.,  {Korista} K.~T.,   {Yakovlev} D.~G.,  1996,
  \mn@doi [\apj] {10.1086/177435}, \href
  {https://ui.adsabs.harvard.edu/abs/1996ApJ...465..487V} {465, 487}

\bibitem[\protect\citeauthoryear{{Vinciguerra} et~al.,}{{Vinciguerra}
  et~al.}{2020}]{Vinciguerra2020}
{Vinciguerra} S.,  et~al., 2020, \mn@doi [\mnras] {10.1093/mnras/staa2177},
  \href {https://ui.adsabs.harvard.edu/abs/2020MNRAS.498.4705V} {498, 4705}

\bibitem[\protect\citeauthoryear{{Walter}, {Lutovinov}, {Bozzo}  \&
  {Tsygankov}}{{Walter} et~al.}{2015}]{Walter2015}
{Walter} R.,  {Lutovinov} A.~A.,  {Bozzo} E.,   {Tsygankov} S.~S.,  2015,
  \mn@doi [Astronomy and Astrophysics Review] {10.1007/s00159-015-0082-6},
  \href {https://ui.adsabs.harvard.edu/abs/2015A&ARv..23....2W} {23, 2}

\bibitem[\protect\citeauthoryear{{Wilms}, {Allen}  \& {McCray}}{{Wilms}
  et~al.}{2000}]{Wilms2000}
{Wilms} J.,  {Allen} A.,   {McCray} R.,  2000, \mn@doi [\apj] {10.1086/317016},
  \href {http://adsabs.harvard.edu/abs/2000ApJ...542..914W} {542, 914}

\bibitem[\protect\citeauthoryear{{Yang}, {Laycock}, {Christodoulou},
  {Fingerman}, {Coe}  \& {Drake}}{{Yang}
  et~al.}{2017}]{Yang2017ApJ...839..119Y}
{Yang} J.,  {Laycock} S.~G.~T.,  {Christodoulou} D.~M.,  {Fingerman} S.,  {Coe}
  M.~J.,   {Drake} J.~J.,  2017, \mn@doi [\apj] {10.3847/1538-4357/aa6898},
  \href {https://ui.adsabs.harvard.edu/abs/2017ApJ...839..119Y} {839, 119}

\bibitem[\protect\citeauthoryear{{Zolotukhin}, {Bachetti}, {Sartore},
  {Chilingarian}  \& {Webb}}{{Zolotukhin} et~al.}{2017}]{2017ApJ...839..125Z}
{Zolotukhin} I.~Y.,  {Bachetti} M.,  {Sartore} N.,  {Chilingarian} I.~V.,
  {Webb} N.~A.,  2017, \mn@doi [\apj] {10.3847/1538-4357/aa689d}, \href
  {https://ui.adsabs.harvard.edu/abs/2017ApJ...839..125Z} {839, 125}

\makeatother
\end{thebibliography}

\bsp	
\label{lastpage}

\end{document}